\documentclass[
  reprint,           % final two-column layout
  pra,               % journal = Physical Review A
  nofootinbib,       % no footnotes in bibliography
  amsmath,amssymb,   % AMS math packages
  aps,               % APS styling
  floatfix           % handle floating figures better
]{revtex4-2}

\usepackage{graphicx}% Include figure files
\usepackage{dcolumn}% Align table columns on decimal point
\usepackage{bm}% bold math
\usepackage{soul}
\usepackage{hyperref}
\usepackage{xcolor}
\usepackage{hyperref}
\graphicspath{{./Images/}}
\begin{document}
\preprint{APS/123-QED}
\title{Long-distance cascaded fluorescence of cold Cesium atoms coupled to an optical nanofiber}
%\title{Long-distance feedback to cold atoms coupled to an optical nanofiber}
\author{M.  Sadeghi}
\email{msad106@aucklanduni.ac.nz}
\author{W. Crump}
%\email{wayne.crump@auckland.ac.nz}
\author{S.  Parkins}%
 %\email{s.parkins@auckland.ac.nz}
 \author{M.D. Hoogerland}
 \email{m.hoogerland@auckland.ac.nz}
\affiliation{%
Department of Physics and Dodd-Walls Centre for Photonic and Quantum Technologies,
University of Auckland,  Private Bag 92019,  Auckland,  New Zealand} 
\begin{abstract}
We demonstrate the first experimental realization of cascaded resonance fluorescence over a 64-meter propagation delay time between two spatially and temporally independent ensembles of laser-cooled Cesium atoms coupled to an optical nanofiber. Spontaneously emitted photons from a strongly driven first ensemble are guided through a standard fiber, reflected by a fiber Bragg grating mirror, and interact with a second ensemble, producing a unidirectional two-node cascaded system. The cascaded fluorescence spectrum is broadened and blue-shifted relative to the original fluorescence spectrum. Our simple model reproduces the power broadening and the cascaded fluorescence spectrum, as well as the ratio of cascaded to original photon flux, giving insight into non-Markovian dynamics. Our results establish the longest-distance one-way cascaded atom-photon interface reported to date, providing a stepping stone towards a fiber-based platform for quantum networking.
\end{abstract}
\maketitle
\section{\label{Sec:Introduction}Introduction}
%\paragraph{\textbf{Cold atoms for Quantum information over long distances:}}
Neutral atoms cooled to micro-Kelvin temperatures provide a nearly ideal two-level system because of 
%characterized by 
their coherence times and precise laser control~\citep{phillips1998nobel,chu1998nobel}. They have already enabled ground-breaking demonstrations of long-distance entanglement distribution through quantum repeaters~\cite{duan2001long} and fiber-based quantum networking prototypes~\cite{kimble2008quantum}. A current frontier is to interface these atoms directly to optical nanofibers, where strong evanescent coupling can turn spontaneous emission into a resource for quantum links~\cite{vetsch2010optical_1}. Their properties also make them ideal for exploring memory effects in reservoir dynamics, particularly in cascaded configurations~\citep{carmichael1993quantum,gardiner1993driving}, and for investigating non-Markovian dynamics driven by spectral correlations~\citep{breuer2016colloquium}.

%\paragraph{\textbf{Cascaded system:}}
The generation and study of cascaded nonclassical light is vital to a complete understanding of open quantum systems~\citep{carmichael1993quantum,gardiner1993driving}. An open quantum system interacts with an external quantum field, known as a bath, reservoir, or environment, which considerably influences the dynamics and coherence properties of the system~\citep{breuer2016colloquium}. Understanding system-bath interactions is fundamental for long-distance quantum communication and quantum networking~\citep{duan2000quantum,kimble2008quantum}. In general, the dynamics of open quantum systems can be categorized into two main regimes: Markovian and non-Markovian~\cite{gardiner2004quantum}.

%\paragraph{\textbf{Markovian and non-Markovian regime:}}
In the Markovian regime, the bath does not retain memory of the system's past state over significant timescales~\citep{kraus2008preparation, lindblad1976generators}. The evolution depends only on the present state, as the environment immediately loses information about past system states. Although this assumption simplifies the theoretical description of quantum dynamics~\cite{gardiner1992wave,parkins1993spectral} and is a reliable approximation for many experiments~\cite{gardiner2015quantum}, it does not adequately reflect more complex quantum phenomena in the real world, such as cascaded quantum systems~\cite{carmichael1993quantum,ulhaq2012cascaded, clark2003unconditional, liedl2024observation,suarez2025chiral} or quantum feedback~\cite{wiseman1993quantum,whalen2017open,nemet2016enhanced,nemet2019stabilizing}. 

In contrast, non-Markovian effects emerge when the reservoir retains memory of past states of the quantum system for a non-negligible time~\cite{lambropoulos2000fundamental,de2017dynamics,breuer2016colloquium}. For example, if emitted photons from an ensemble of atoms reflect off a mirror and return after a delay comparable to the atomic lifetime, the reservoir retains and reintroduces information about the earlier state of the ensemble~\cite{eschner2001light,dubin2007photon}. This possible time-delayed self-interaction, namely coherent time-delayed feedback (CTDF), can substantially impact the system evolution~\cite{de2017dynamics,parkins1988effect,nemet2019time}.

%\paragraph{\textbf{Our system:}}
In this paper, we consider a quantum system of Cesium atoms coupled to a distant mirror at a separation corresponding to a time delay in light propagation that is significantly longer than the atomic excited-state lifetime (in our experiment, over ten times the $\sim 30.4$ ns lifetime of Cesium~\cite{steck2003cesium}). In this case, emitting atoms reach a steady state before photons reflected from the mirror return. As a result, this quantum system cannot be treated as a single-ensemble case, 
%(single-ensemble feedback drive) 
such as the one in which atoms interact with their own emitted photons after a delay comparable to the atomic lifetime (such as in the conventional CTDF scenario described above). Instead, this configuration can be described as two separate ensembles (a two-ensemble unidirectional drive). The substantial delay between the emission of the first ensemble and its reflection establishes a cascaded quantum system interaction, where emission from the first ensemble unidirectionally
%\footnote{Unidirectional coupling denotes emission propagating in one direction to drive subsequent systems, e.g., from a first atomic ensemble to a second, without reverse interaction, ensuring a one-way propagation of excitations, applicable to networks of cascaded quantum systems.} 
drives the second ensemble, as originally theoretically considered by Carmichael~\cite{carmichael1993quantum} and Gardiner~\cite{gardiner1993driving}. Although the photon round-trip delay far exceeds the memory timescale of the atoms, effectively rendering the delay memoryless, in this quantum system a non-Markovian memory effect arises from the finite bandwidth and finite correlation time of the emission spectrum~\citep{breuer2016colloquium}.

%\paragraph{\textbf{ONFs:}}
%Experimental studies of quantum coherence and memory effects require controlled and well-characterized interactions between quantum emitters and their environment. 
We employ optical nanofibers (ONFs), which offer an excellent platform by enabling strong coupling to guided modes through their evanescent field~\cite{vetsch2010optical_1,gouraud2015demonstration,goban2015strong,nayak2018nanofiber}. Previous studies involving ONFs typically measured absorption spectra to study atom-light interactions~\cite{ruddell2017collective} and non-Markovian memory effects~\cite{lechner2023light}. More recently, emission properties have been studied by exciting atoms around the ONF using a free-space side probe beam~\cite{solano2017super,corzo2019waveguide,solano2019alignment}, which illuminates atoms externally, often perpendicular to the fiber axis, enabling emission into the guided mode of the nanofiber. These studies provide insights into quantum coherence~\cite{nayak2007optical,solano2019alignment,das2010measurement} and non-Markovian memory effects in emission~\cite{solano2017super,sinha2020non,ferreira2021collapse}.

%\paragraph{\textbf{Papaer overview:}}
%In this paper, 
We present the first experimental realization of cascaded fluorescence from cold Cesium atoms near an ONF, where a correlated noisy field
%\footnote{Correlated noise refers to the resonance fluorescence spectrum driving the cascaded interaction.} from the first ensemble fluorescence, with a Mollow spectrum~\cite{mollow1969power}\footnote{The Mollow structure refers to the emission spectrum of a two-level atom under strong driving regime, featuring a central peak and two sidebands due to dressed state splitting.}, 
unidirectionally drives a second ensemble. Unlike CTDF, where feedback photons reintroduce past states, non-Markovianity here arises from the nontrivial spectrum of the noisy bath. Such a finite-bandwidth drive contains non-zero two-time correlations. In our case, the drive for the second ensemble is given by resonance fluorescence, which has a correlation function that decays on a timescale comparable
to the atomic lifetime and, therefore, feeds information from the past back
into the system, yielding non-Markovian dynamics. Because the driving light exhibits sidebands and altered photon statistics beyond simple monochromatic or thermal fields, the second ensemble sees the noise as a non-Markovian drive.
%\footnote{In the white-noise limit of infinite bandwidth, these correlations vanish, and the drive becomes Markovian.}.
In this case, we cannot simply model the environment as a memoryless Markovian reservoir or with Markovian Bloch equations~\cite{carmichael2007statistical}. Instead, we must account for the spectral correlations of the light that is cascaded into the second ensemble~\cite{breuer2016colloquium}. We study the original fluorescence and cascaded fluorescence spectra, revealing spectral features such as absorption broadening. Fitting the cascaded spectra, taking into account the Mollow spectrum of the initial emission, provides good agreement with the experimental data and allows us to quantify the effects of strong driving and long-distance cascading observed in our experiment.

\section{\textbf{Experimental configuration}}
\begin{figure}
\centering
	\includegraphics[width=1 \linewidth]{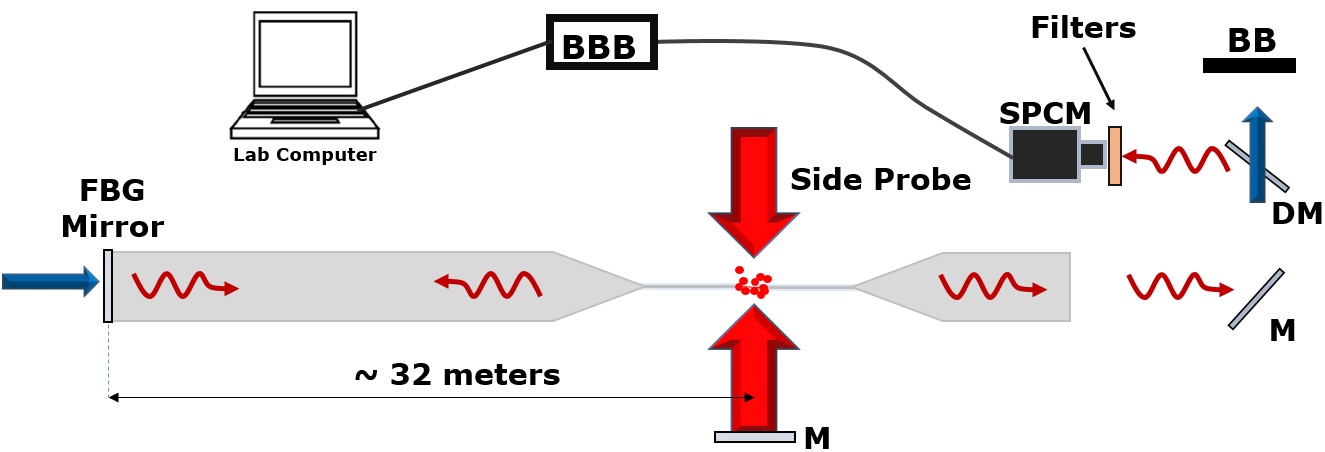} 
	\caption{Schematic of the experimental setup. The FBG mirror is approximately 32 meters away from the atoms, with a reflectance of 99\%. The components are filters, a beam blocker (BB), a single-photon counting module (SPCM), a Beaglebone black (BBB), a mirror (M), and a dichroic mirror (DM). The probe is perpendicular to the nanofiber waist.}
	\label{article_setup}
\end{figure}
%\paragraph{\textbf{Experiment setup:}}
The diagram in Fig.~\ref{article_setup} presents a schematic of the experimental setup used in this study.  A Magneto-Optical Trap (MOT)~\cite{chu1991laser,ruddell2017calorimetry} creates a cold cloud of approximately 5 million atoms at a temperature of around $150\,\mu$K surrounding the waist of an ONF, which has a diameter of approximately $380$~nm.  At typical MOT densities in our experimental setup, we estimate fewer than ten atoms within the evanescent field of the ONF. The MOT setup is similar to the one used in Ref.~\cite{ruddell2017collective}. At the center of the chamber, the waist of the ONF intersects with the cloud of cold Cesium atoms produced by the MOT. The Cesium cloud is loaded from a background gas generated by a ${}^{133}\text{Cs}$ dispenser. 

%\paragraph{\textbf{ONF fabrication:}}
The ONF was fabricated in-house using the flame-brush technique, starting from a standard single-mode optical fiber. During fabrication, the fiber is simultaneously heated and stretched, creating an adiabatic transition between the standard and stretched portions of the fiber~\cite{birks1992shape,karapetyan2011optical,nagai2014ultra, ruddell2017calorimetry}. The ONF is single-mode at the $D_{2}$ resonant wavelength of 852 nm and is placed inside an ultra-high-vacuum chamber. 

%\paragraph{\textbf{685 laser:}}
Atoms very close to the ONF adhere to its surface due to van der Waals forces~\cite{nayak2012spectroscopy}, resulting in a Cesium coating that forms after a few minutes of exposure. This coating reduces light transmission to less than 40\%. To suppress such a surface coating, a 685 nm red-detuned laser beam with 0.3 mW power is applied during MOT loading. The repulsive beam is switched off for 500 $\mu$s before the MOT beams are extinguished, ensuring that the red beam is absent throughout the subsequent probe sequence. This timing allows atoms to approach the nanofiber surface for probing, which also allows surface-induced interactions~\cite{passerat2006laser,nayak2007optical,kien2007phonon,kien2007optical,le2007spontaneous,das2010measurement,patterson2018spectral}.  

%\paragraph{\textbf{exciting laser beam:}}
The MOT atoms surrounding the ONF are excited using probe pulses close to resonant with the $F=4 \longrightarrow F^{\prime}=5$ transition of the ${}^{133}\text{Cs}$ $D_{2}$ line. After the MOT cooling beams are switched off, we leave 10 $\mu$W of the repump laser on during the probe sequence to maintain the atomic population in the $F=4$ ground state, ensuring consistent excitation conditions. An Acousto-Optic Modulator controls the exciting beam's power and frequency in a double-pass configuration. The beam is coupled into a polarization-maintaining, single-mode optical fiber. The fiber output is collimated to a beam diameter ($1/e^2$) of 1.5~mm, and forms a side excitation beam that propagates perpendicularly to the ONF waist in a standing wave configuration. The standing wave arrangement reduces momentum transfer from the exciting laser to the atoms, increasing emission into the ONF. 
%The collimated side probe beam has a $ 1/e^2$ full width of approximately 1.5~mm.

%\paragraph{\textbf{exciting laser pulses:}}
The radio frequency (RF) signal to drive the Acousto-Optic Modulator is derived from a digital signal generator around 80~MHz. The RF amplitude is simultaneously adjusted to obtain a constant beam power. The RF is also switched by an Arbitrary Function Generator (Tektronix 3032) using an RF switch. A train of 150~ns long pulses, repeated every 600~ns, was programmed into the Function Generator and triggered by the experiment computer for each experiment.

%\paragraph{\textbf{original and cascaded photons:}}
Spontaneous emission photons couple into the ONF-guided mode for each laser pulse in both directions. The linearly polarized side probe beam excites the $F=4 \to F'=5$ transition of $^{133}$Cs, involving multiple Zeeman sublevels. Due to the near-uniform Clebsch-Gordan coefficients for these sublevels under the near-zero magnetic field in our experiment, the emission is approximately isotropic, with 50\% directed toward the Single Photon Counting Module (Perkin-Elmer SPCM-AQRH-3X-W3) and 50\% toward the fiber Bragg grating (FBG) mirror. This symmetry allows us to treat the atoms as an effective two-level system for linewidth and saturation analysis. Photons propagating along a short length of single-mode fiber to the SPCM are termed original photons (or original fluorescence). Meanwhile, the original photons emitted in the opposite direction travel through a longer length of single-mode fiber to the FBG mirror, reflect, and interact with a second ensemble of atoms that have reached the ground state during the propagation time. These reflected photons are attenuated by fiber losses and the second ensemble before detection at the SPCM. We refer to these photons as cascaded photons or cascaded fluorescence. 

%\paragraph{\textbf{Photon detection:}}
The SPCM output pulses are time-tagged by the Programmable Realtime Unit on a BeagleBone Black single-board computer with a time resolution of 5~ns. The arrival times of all photons from one experiment run, consisting of 2000 exciting laser pulses, are transferred to the laboratory computer after each experimental cycle.

\begin{figure}
\includegraphics[scale=0.5]{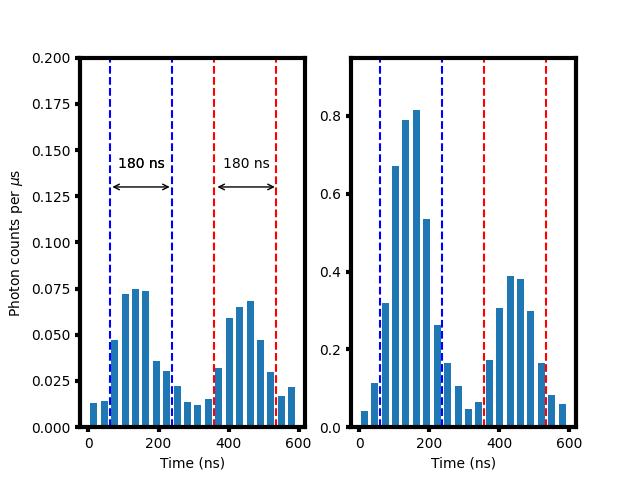}
\caption{Photon arrival histogram for an excitation laser power of 40 $\mu$W, which corresponds to $s_0=I/I_0=0.4$, for two values of the laser detuning: $\Delta=30$~MHz (left) and $\Delta=0$~MHz (right). The first peak represents photons that have been emitted into the ONF towards the detector. The second peak represents photons that have been emitted towards the FBG mirror, reflected, interacted with the ground state atoms in the MOT before being detected. In the data analysis, photon counts from emission and feedback photons were considered within the 180 ns time interval indicated by the horizontal arrows. \label{fig:histogram}}
\end{figure}

%\paragraph{\textbf{Experimental sequence:}}
For each experiment, the MOT is filled with atoms at a MOT laser detuning of 12~MHz. The MOT lasers are then switched to a detuning of 22~MHz for additional cooling for 0.5~ms. Subsequently, the MOT lasers are extinguished, and the excitation laser pulse train is triggered. The experiment is run 120 times for each detuning and excitation power, for all the data presented in this paper. In Fig.~\ref{fig:histogram}, we show histograms of the arrival times of photons on the detector, with the ``cascaded'' photons arriving around 310~ns later than the ``original'' photons. For the results in the left pane, the exciting laser detuning is 30~MHz, large compared to the natural linewidth of the transition.
The cascaded photon counts are roughly 90\% of the original counts, reflecting losses along this path, such as attenuation by the fiber and imperfect reflection from the FBG mirror. However, when the exciting laser is in resonance with the atoms around the ONF, the right pane in Fig.~\ref{fig:histogram},  the cascaded photons are attenuated significantly, caused by cascaded absorption of the original fluorescence by the second ensemble around the ONF.
 
%\paragraph{\textbf{Data collection:}}
As the exciting probe laser is heating the atoms in the cloud, we found that the count rate diminishes after a number of excitation pulses, depending on the probe laser intensity and detuning. Therefore, we have chosen to limit the number of detected photons to 1500 for each atom cloud in post-processing, as this number of detected photons is a good indicator of the degree of heating in the cloud. Given the time when the last photon is detected, we establish a count rate, also known as photon flux.

\section{\textbf{Experimental Results}}
\label{results}
%\paragraph{\textbf{Intro for results:}}
%In this section, we present the main results of our experiment, focusing on the spontaneously emitted photons of cold Cesium atoms coupled to an ONF for various excitation intensities and detunings and considering the \textcolor{blue}{spectrum} of the cascaded fluorescence arising from the interaction of reflected \textcolor{blue}{original} fluorescence photons with the second ensemble. 

\begin{figure}
    \centering
    \includegraphics[width=\linewidth]{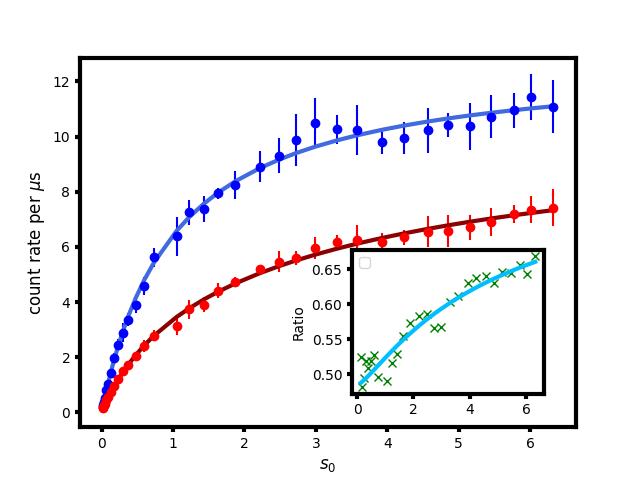}
    \caption{The photon count rate for the original (blue dots) and cascaded (red dots) fluorescence as a function of the saturation parameter $s_0$. In the inset, the ratio of the count rates is shown. The laser detuning $\Delta=0$. The points represent the mean photon count rates, and the error bars show the standard deviation across four separate experiments.}
    \label{fig:saturation}
\end{figure}

%\paragraph{\textbf{saturation analysis:}}
We first examine the saturation behavior of the system, followed by an analysis of the cascaded spectrum using the model discussed below. Fig.~\ref{fig:saturation} presents the count rates of the original fluorescence and the cascaded fluorescence as a function of the laser intensity for on-resonance excitation. We scale the intensity such that half of the maximum count rate for the original fluorescence is reached at $s_0=1$, following the textbook two-level relation~\cite{metcalf1999laser} for the excited state population $n_e$, 
\begin{equation}
    n_e=\frac{s_0}{2(1+s_0)} \label{eq:s0},
\end{equation}
where $s_0=I/I_0$, $I$ is the laser intensity, and $I_0$ is the saturation intensity. Because a few atoms interacting with the ONF mode result in negligible cooperative dipole-dipole effects~\cite{ott2013cooperative}, the ensemble can be treated as an effective single two-level system. This scaling yields a saturation power of 121~$\mu$W, which agrees with our estimates of the laser beam size. The blue curve in the graph corresponds to Eq.~(\ref{eq:s0}).
 In the inset of Fig.~\ref{fig:saturation}, we show the ratio of the cascaded and original fluorescence rate. The relative amount of cascaded absorption decreases as the exciting laser power increases.

%\paragraph{\textbf{Mollow spectrum:}}
The atoms around the nanofiber are excited by a laser beam that may exceed the saturation intensity. In this strong driving regime (\(\Omega = \Gamma \sqrt{s/2} > \Gamma/4\)), the spectrum of the emission into the fiber exhibits the triplet structure introduced by Mollow~\cite{mollow1969power}. This fluorescence spectrum can be expressed as~\cite{ortiz2019mollow},
\begin{widetext}
\begin{equation}
% \label{}
    g_{sc}(\omega)=\frac{s}{(2+s)^2}\delta(\omega)
    + \frac{s_0}{8\pi\Gamma}\frac{s}{1+s} 
    \times \frac{1+s_0/4+(\omega/\Gamma)^2}{\left[ \frac{1}{4} +\frac{s_0}{4}+\left(\frac{\Delta}{\Gamma}\right)^2 -2 \left(\frac{\omega}{\Gamma}\right)^2\right]^2 + \left(\frac{\omega}{\Gamma}\right)\left[ \frac{5}{4}+ \frac{s_0}{2}+\left(\frac{\Delta}{\Gamma}\right)^2 -\left(\frac{\omega}{\Gamma}\right)^2 \right]^2 },
    \label{eq:mollow}
\end{equation}
\end{widetext}
where the first term represents elastic scattering at frequency $\omega$ relative to the laser frequency, $\omega$, and the second term is inelastic scattering with $\Delta$ the laser detuning and $\Gamma$ the natural decay rate. The ratio of the inelastic and elastic scattering power comes out as the frequency-dependent saturation parameter $s$, defined by~\cite{ortiz2019mollow}
\begin{equation}
s(\Delta)=\frac{s_0}{1+4(\Delta/\Gamma)^2}.
\end{equation}

% \paragraph{Fitting the indirect photons}
To obtain the fitted curve for the cascaded fluorescence in Fig.~\ref{fig:saturation} (red curve), we first take the Mollow spectrum at each excitation intensity by summing its inelastic and elastic contributions (see Eq.~\eqref{eq:mollow}). We then normalize this spectrum by the measured number of original photons at that same intensity, reflecting that, on average, the total number of emitted photons is the same in both directions. The original photons with this spectrum interact with the ground-state atoms in the evanescent field of the guided mode, where they experience additional, frequency-dependent attenuation described by the Beer-Lambert law,
\begin{equation}
I = I_0 e^{-\alpha L},
\label{eq:4}
\end{equation}
where $I_0$ is the intensity of the reflected original photons from the FBG mirror, $\alpha$ is the optical depth, and $L$ represents the frequency-dependent Lorentzian absorption profile. Finally, we integrate the Mollow spectrum, weighted by the Beer–Lambert Lorentzian filter, over a scattered-photon frequency offset
$\omega$ (from $-10\Gamma$ to $+10\Gamma$) to obtain the simulated cascaded
photon count (red curve in Fig.~\ref{fig:saturation}). Using the width and depth of the cascaded absorption spectrum as parameters, we obtain a best fit with a width of $(6.7\pm 0.6)$~MHz and an optical depth of $ (0.85\pm 0.04) $. Note that the changing ratio of count rates is well reproduced by this model for higher excitation laser powers. The trend indicates that at higher excitation laser intensities, the Mollow sidebands shift out of resonance to the absorption profile of the second ensemble, resulting in reduced cascaded absorption, as observed in the inset of Fig.~\ref{fig:saturation}.

\begin{figure}
\centering
\includegraphics[width=\linewidth]{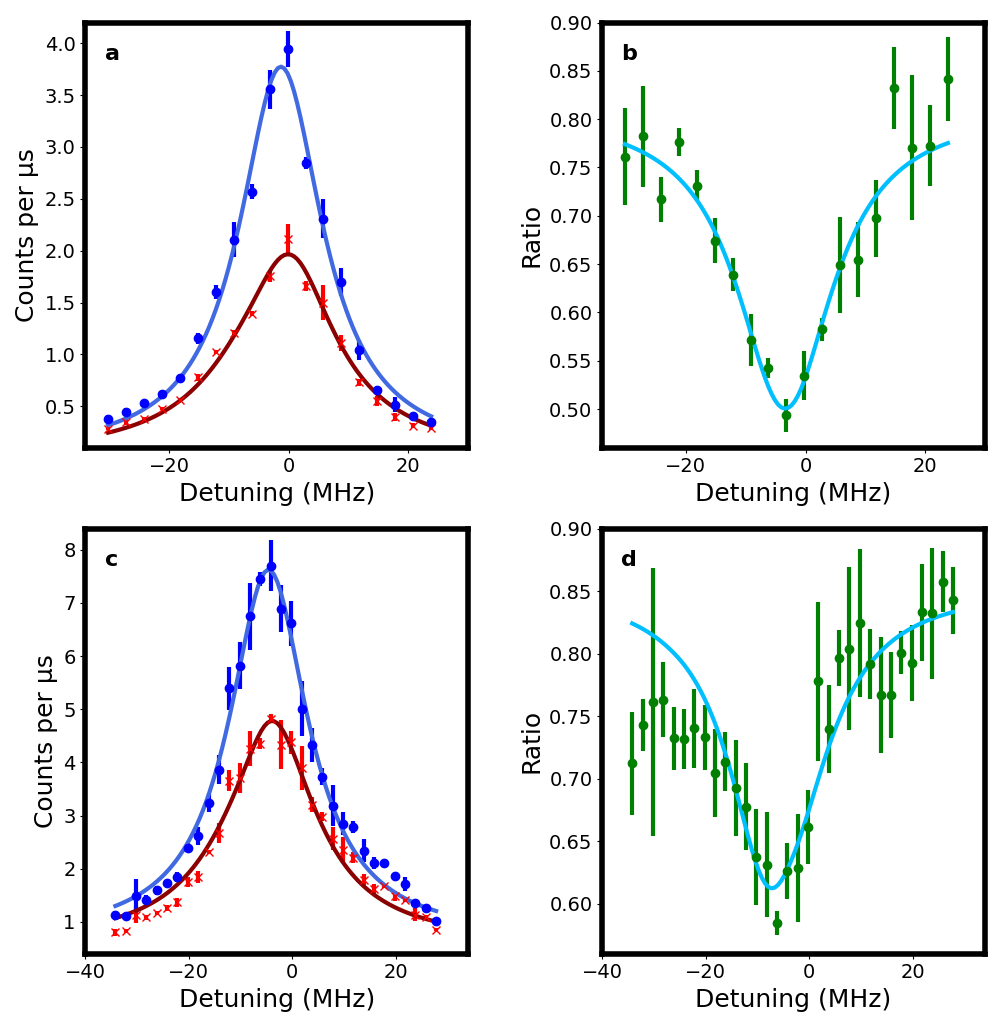}
\caption{(a, c) The photon count rate for original fluorescence spectrum (blue dots) and cascaded fluorescence spectrum (red crosses) as a function of detuning for excitation powers of (a) 40~$\mu$W and (c) 300~$\mu$W, corresponding to saturation parameters $s_0 = I/I_0 = 0.4$ and $s_0 = 2.5$, respectively. Data points represent the mean values, and the error bars indicate the standard deviation from three independent datasets. The blue line for the original fluorescence spectrum is a Lorentzian fit. (b, d) The ratio of the cascaded to original fluorescence spectra as a function of detuning for the same data. Error bars for the ratio represent the uncertainties calculated using standard error propagation, based on three repetitions of the experiment. The fit curves are modeled as described in the text.}
\label{Width_EF_expert}
\end{figure}

%\paragraph{\textbf{Detuning:}}
 In a second series of measurements, we scan the probe detuning and record the count rates for both the original and cascaded fluorescence spectra, as shown in Fig.~\ref{Width_EF_expert}(a, c). The fitted Lorentzian width of the original fluorescence spectrum at the lowest power ($s_0=0.4$) is 16 MHz, approximately three times the natural width of the transition, which is $\sim$~5.2~MHz. This broadening is similar to that observed in previous ONF spectroscopy studies~\cite{nayak2007optical,nayak2012spectroscopy,patterson2018spectral} and is attributed to surface-induced van der Waals potentials that vary across the first ensemble. Nonetheless, for the range of detunings and excitation intensities presented in this paper, both original and cascaded fluorescence spectra retain their Lorentzian lineshapes~\cite{kien2007optical,nayak2012spectroscopy,patterson2018spectral}.

\begin{figure}
\centering
\includegraphics[width=\linewidth]{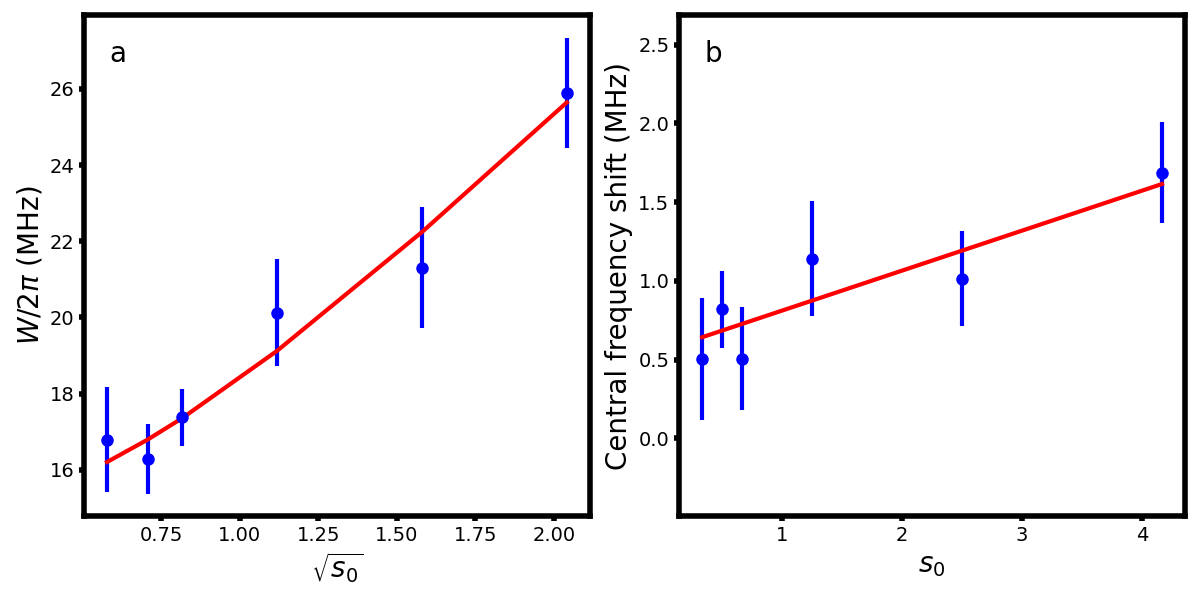}
\caption{
(a) The original fluorescence spectrum linewidth as a function of the square root of the saturation parameter, showing an increase in linewidth with excitation power. The total linewidth is the sum of the natural linewidth and additional broadening.
(b) The central frequency shift between original and cascaded fluorescence as a function of saturation parameter. The shift increases with the excitation power. The error bars indicate uncertainties.
}
\label{fig:linewidth}
\end{figure}

Increasing the free-space probe power redshifts the center frequency of the original fluorescence (blue curves in Fig.~\ref{Width_EF_expert}(a,c)). The original resonance fluorescence becomes approximately 4.5 MHz redshifted as the probe power increases from $40\;\mu\text{W}$ to $500\;\mu\text{W}$. We attribute this shift to a growing contribution from atoms located closer to the ONF surface, where the van der Waals potential redshifts the transition frequency~\cite{patterson2018spectral,kien2007optical}. Two mechanisms enhance the contribution of these redshifted atoms as the probe power increases: (i) the increased Rabi bandwidth excites atoms whose resonance is already surface-shifted, and (ii) the increasing radiation pressure redistributes the atoms around the ONF, such that the near-surface population during the probe pulse is increased. At the same time, both mechanisms can contribute to the observed redshifts in the original resonance fluorescence. The latter, in the case where the side probe is off, may also contribute to the cascaded fluorescence being blue-shifted, as illustrated in Fig.~\ref{fig:linewidth}(b).

We then model the cascaded fluorescence photons as follows: (1) For each excitation laser detuning $\Delta$, the number of photons emitted towards the FBG mirror is the same as the number originally emitted towards the detector, $n_p(\Delta)$. (2) The spectrum of the emitted photons is given by Eq.~(\ref{eq:mollow}) but normalized with the measured original fluorescence. (3) After reflecting off the FBG  mirror, these photons may be absorbed by the atoms, following the Beer-Lambert law and a Lorentzian absorption profile, with a width given by the width of the original emission spectrum at low power but with a frequency shift as a free parameter. (4) The observed cascaded photon count is obtained by integrating the absorbed original photon spectrum over the scattered-photon frequency offset $\omega$ (from $-10\Gamma$ to $+10\Gamma$), as described in the power-scan procedure above. Red curves in Fig.~\ref{Width_EF_expert}(a, c) show the fit to the measured cascaded photon flux. At higher power, the spectra exhibit small asymmetric tails, resulting from power-dependent heating and surface-induced interactions~\cite{patterson2018spectral}.

Furthermore, we take the ratio of the cascaded and original photons as a function of detuning, as shown in Fig.~\ref{Width_EF_expert}(b, d). This ratio reflects the frequency-dependent attenuation of the original fluorescence spectra. As a consequence of this attenuation, the cascaded fluorescence spectra are, on average, 5 MHz broader than the original fluorescence spectra. We observe that as the probe intensity increases, the central dip of the ratio decreases. This reduction in the dip, again, is attributed to the fact that fewer cascaded absorption events of the reflected original photons occur as the driving laser intensity increases, consistent with the results displayed in the inset of Fig.~\ref{fig:saturation}. The fitted curves in Fig.~\ref{Width_EF_expert}(b, d) are from the ratio of the fitted curves in Fig.~\ref{Width_EF_expert}(a, c). The large error bars in the ratio arise from increased relative uncertainties as the detuning increases, where both counts approach similar values near background levels, thereby raising relative uncertainties through error propagation. Additionally, the slight asymmetric tails observed in Figs.~\ref{Width_EF_expert}(a, c), particularly at higher power, and the additional frequency shift described in Fig.~\ref{fig:linewidth}(b), contribute to the large uncertainties in the ratio.

%\paragraph{\textbf{Power broadening.}}
For a driven two-level atom, the fluorescence linewidth increases with the optical Rabi frequency \(\Omega\)~\cite{metcalf1999laser}.  To verify this power-broadening behavior, we recorded the original excitation spectra at several probe intensities. We extracted the full width at half maximum of the original fluorescence for each case. Figure~\ref{fig:linewidth}(a) shows these linewidths as a function of
\(\sqrt{s_0}\), where \(s_0=I/I_0\) is the on-resonance saturation
parameter.
The data follows the expected law,
\begin{equation}
W(s_0)=\Gamma \sqrt{s_0+1}+\Gamma_0 ,
\end{equation}
in which the \(\Gamma\sqrt{s_0+1}\) term represents power broadening, $\Gamma$ is the natural atomic decay rate of the system and \(\Gamma_0\) collects additional broadening from surface interactions and other
residual dephasing effects. From the fitting, we obtain $\Gamma$ = (6.45 ± 1.17) MHz and $\Gamma_0$ = (8.44 ± 0.80) MHz, implying that the original fluorescence linewidth of the system closely matches the cascaded absorption linewidth estimated previously.

%\paragraph{\textbf{Freq shift with power:}}
The relative center of the cascaded fluorescence spectrum, analyzed in Fig.~\ref{fig:linewidth}(b), reveals an additional small blue shift of 0.5 to 1.5 MHz relative to the original fluorescence, extracted via Lorentzian fits across all powers. This shift increases roughly linearly with excitation power (slope 0.25 ± 0.06 MHz), reflecting the second ensemble redistribution during the 310 ns propagation delay. During the photon propagation delay, atoms in the first ensemble can experience a power-dependent distribution within the van der Waals interaction range~\cite{das2010measurement,solano2019alignment}, potentially altering their coupling to the ONF and further shifting the resonance. Unlike the first ensemble, where the probe intensity drives the shift, the second ensemble's shift arises from increased temperature and cascaded absorption. Higher probe intensities induce greater heating and alter the excitation of the atoms by the reflected original photons.

\section{Conclusion and outlook}
We have experimentally demonstrated a two-node, cascaded interface in which photons emitted by a strongly driven ensemble of cold \textsuperscript{133}Cs atoms are routed through an optical nanofiber and interact with a second, ground-state ensemble. As the probe intensity increases, it excites a broader range of atoms in the first ensemble, including those closer to the ONF surface, which results in a redshift of the first ensemble and a corresponding redshift of the original fluorescence. The power-dependence redshift also influences the atoms during the 310 ns propagation delay time when the side probe is off. The increased side probe power leads to increased heating, which further redistributes and redshifts the atoms as they form the second ensemble. The result of this is that the cascaded fluorescence moves toward the bare resonance and therefore is slightly blue-shifted relative to the red-shifted original fluorescence.

A power-broadened Mollow triplet filtered by Beer-Lambert absorption shows the power dependence of the cascaded spectrum, demonstrating that the extra broadening is due to cascaded absorption within the second ensemble. The extracted cascaded absorption linewidth of the second ensemble \((6.7\pm0.6)\,\mathrm{MHz}\) is comparable to the original emission linewidth of the first ensemble \((6.45\pm1.17)\,\mathrm{MHz}\), confirming that a finite-bandwidth drive gives rise to non-Markovian memory effects over distances far exceeding the excited-state lifetime. 

Future work will investigate cascaded atom-photon interactions in the regime of enhanced coherence \cite{goban2015strong,solano2017super,liedl2024observation}. Optimizing the ONF for lower loss \cite{ruddell2020ultra} and trapping atoms near the ONF surface using a repump beam~\cite{solano2017super,patterson2018spectral} allows us to drive the system from the weak-collective limit into the super-radiant regime~\cite{dicke1954coherence,solano2017super}. Photon correlation measurements could explore whether absorption-induced coherence generates collective super-radiant bursts in cascaded configurations~\cite{liedl2024observation} and multi-ensemble networks~\cite{tebbenjohanns2024predicting}. A transfer matrix model~\cite{nemet2020transfer} based on the optical Bloch equations can be extended to simulate such cascaded systems, incorporating dephasing mechanisms in emission, as well as cascaded linewidth and spatial system geometry, such as propagation efficiency and atom-ONF coupling efficiency of the system.

In addition, reducing the nanofiber-mirror distance to shorten the photon round-trip time toward the Cesium lifetime will let us explore the
crossover from the unidirectional cascaded limit~\cite{carmichael1993quantum} to a coherent time-delayed feedback regime~\cite{pichler2016photonic,nemet2019time}. Investigating this crossover will provide new insights into how memory effects influence emission spectra, temporal photon correlations, and photon flow in open quantum systems.

\bibliography{Article}

\begin{thebibliography}{63}%
\makeatletter
\providecommand \@ifxundefined [1]{%
 \@ifx{#1\undefined}
}%
\providecommand \@ifnum [1]{%
 \ifnum #1\expandafter \@firstoftwo
 \else \expandafter \@secondoftwo
 \fi
}%
\providecommand \@ifx [1]{%
 \ifx #1\expandafter \@firstoftwo
 \else \expandafter \@secondoftwo
 \fi
}%
\providecommand \natexlab [1]{#1}%
\providecommand \enquote  [1]{``#1''}%
\providecommand \bibnamefont  [1]{#1}%
\providecommand \bibfnamefont [1]{#1}%
\providecommand \citenamefont [1]{#1}%
\providecommand \href@noop [0]{\@secondoftwo}%
\providecommand \href [0]{\begingroup \@sanitize@url \@href}%
\providecommand \@href[1]{\@@startlink{#1}\@@href}%
\providecommand \@@href[1]{\endgroup#1\@@endlink}%
\providecommand \@sanitize@url [0]{\catcode `\\12\catcode `\$12\catcode `\&12\catcode `\#12\catcode `\^12\catcode `\_12\catcode `\%12\relax}%
\providecommand \@@startlink[1]{}%
\providecommand \@@endlink[0]{}%
\providecommand \url  [0]{\begingroup\@sanitize@url \@url }%
\providecommand \@url [1]{\endgroup\@href {#1}{\urlprefix }}%
\providecommand \urlprefix  [0]{URL }%
\providecommand \Eprint [0]{\href }%
\providecommand \doibase [0]{https://doi.org/}%
\providecommand \selectlanguage [0]{\@gobble}%
\providecommand \bibinfo  [0]{\@secondoftwo}%
\providecommand \bibfield  [0]{\@secondoftwo}%
\providecommand \translation [1]{[#1]}%
\providecommand \BibitemOpen [0]{}%
\providecommand \bibitemStop [0]{}%
\providecommand \bibitemNoStop [0]{.\EOS\space}%
\providecommand \EOS [0]{\spacefactor3000\relax}%
\providecommand \BibitemShut  [1]{\csname bibitem#1\endcsname}%
\let\auto@bib@innerbib\@empty
%</preamble>
\bibitem [{\citenamefont {Phillips}(1998)}]{phillips1998nobel}%
  \BibitemOpen
  \bibfield  {author} {\bibinfo {author} {\bibfnamefont {W.~D.}\ \bibnamefont {Phillips}},\ }\bibfield  {title} {\bibinfo {title} {Nobel lecture: Laser cooling and trapping of neutral atoms},\ }\href@noop {} {\bibfield  {journal} {\bibinfo  {journal} {Reviews of Modern Physics}\ }\textbf {\bibinfo {volume} {70}},\ \bibinfo {pages} {721} (\bibinfo {year} {1998})}\BibitemShut {NoStop}%
\bibitem [{\citenamefont {Chu}(1998)}]{chu1998nobel}%
  \BibitemOpen
  \bibfield  {author} {\bibinfo {author} {\bibfnamefont {S.}~\bibnamefont {Chu}},\ }\bibfield  {title} {\bibinfo {title} {Nobel lecture: The manipulation of neutral particles},\ }\href@noop {} {\bibfield  {journal} {\bibinfo  {journal} {Reviews of Modern Physics}\ }\textbf {\bibinfo {volume} {70}},\ \bibinfo {pages} {685} (\bibinfo {year} {1998})}\BibitemShut {NoStop}%
\bibitem [{\citenamefont {Duan}\ \emph {et~al.}(2001)\citenamefont {Duan}, \citenamefont {Lukin}, \citenamefont {Cirac},\ and\ \citenamefont {Zoller}}]{duan2001long}%
  \BibitemOpen
  \bibfield  {author} {\bibinfo {author} {\bibfnamefont {L.-M.}\ \bibnamefont {Duan}}, \bibinfo {author} {\bibfnamefont {M.~D.}\ \bibnamefont {Lukin}}, \bibinfo {author} {\bibfnamefont {J.~I.}\ \bibnamefont {Cirac}},\ and\ \bibinfo {author} {\bibfnamefont {P.}~\bibnamefont {Zoller}},\ }\bibfield  {title} {\bibinfo {title} {Long-distance quantum communication with atomic ensembles and linear optics},\ }\href@noop {} {\bibfield  {journal} {\bibinfo  {journal} {Nature}\ }\textbf {\bibinfo {volume} {414}},\ \bibinfo {pages} {413} (\bibinfo {year} {2001})}\BibitemShut {NoStop}%
\bibitem [{\citenamefont {Kimble}(2008)}]{kimble2008quantum}%
  \BibitemOpen
  \bibfield  {author} {\bibinfo {author} {\bibfnamefont {H.~J.}\ \bibnamefont {Kimble}},\ }\bibfield  {title} {\bibinfo {title} {The quantum internet},\ }\href@noop {} {\bibfield  {journal} {\bibinfo  {journal} {Nature}\ }\textbf {\bibinfo {volume} {453}},\ \bibinfo {pages} {1023} (\bibinfo {year} {2008})}\BibitemShut {NoStop}%
\bibitem [{\citenamefont {Vetsch}\ \emph {et~al.}(2010)\citenamefont {Vetsch}, \citenamefont {Reitz}, \citenamefont {Sagu{\'e}}, \citenamefont {Schmidt}, \citenamefont {Dawkins},\ and\ \citenamefont {Rauschenbeutel}}]{vetsch2010optical_1}%
  \BibitemOpen
  \bibfield  {author} {\bibinfo {author} {\bibfnamefont {E.}~\bibnamefont {Vetsch}}, \bibinfo {author} {\bibfnamefont {D.}~\bibnamefont {Reitz}}, \bibinfo {author} {\bibfnamefont {G.}~\bibnamefont {Sagu{\'e}}}, \bibinfo {author} {\bibfnamefont {R.}~\bibnamefont {Schmidt}}, \bibinfo {author} {\bibfnamefont {S.}~\bibnamefont {Dawkins}},\ and\ \bibinfo {author} {\bibfnamefont {A.}~\bibnamefont {Rauschenbeutel}},\ }\bibfield  {title} {\bibinfo {title} {Optical interface created by laser-cooled atoms trapped in the evanescent field surrounding an optical nanofiber},\ }\href@noop {} {\bibfield  {journal} {\bibinfo  {journal} {Physical Review Letters}\ }\textbf {\bibinfo {volume} {104}},\ \bibinfo {pages} {203603} (\bibinfo {year} {2010})}\BibitemShut {NoStop}%
\bibitem [{\citenamefont {Carmichael}(1993)}]{carmichael1993quantum}%
  \BibitemOpen
  \bibfield  {author} {\bibinfo {author} {\bibfnamefont {H.~J.}\ \bibnamefont {Carmichael}},\ }\bibfield  {title} {\bibinfo {title} {Quantum trajectory theory for cascaded open systems},\ }\href@noop {} {\bibfield  {journal} {\bibinfo  {journal} {Physical review letters}\ }\textbf {\bibinfo {volume} {70}},\ \bibinfo {pages} {2273} (\bibinfo {year} {1993})}\BibitemShut {NoStop}%
\bibitem [{\citenamefont {Gardiner}(1993)}]{gardiner1993driving}%
  \BibitemOpen
  \bibfield  {author} {\bibinfo {author} {\bibfnamefont {C.}~\bibnamefont {Gardiner}},\ }\bibfield  {title} {\bibinfo {title} {Driving a quantum system with the output field from another driven quantum system},\ }\href@noop {} {\bibfield  {journal} {\bibinfo  {journal} {Physical review letters}\ }\textbf {\bibinfo {volume} {70}},\ \bibinfo {pages} {2269} (\bibinfo {year} {1993})}\BibitemShut {NoStop}%
\bibitem [{\citenamefont {Breuer}\ \emph {et~al.}(2016)\citenamefont {Breuer}, \citenamefont {Laine}, \citenamefont {Piilo},\ and\ \citenamefont {Vacchini}}]{breuer2016colloquium}%
  \BibitemOpen
  \bibfield  {author} {\bibinfo {author} {\bibfnamefont {H.-P.}\ \bibnamefont {Breuer}}, \bibinfo {author} {\bibfnamefont {E.-M.}\ \bibnamefont {Laine}}, \bibinfo {author} {\bibfnamefont {J.}~\bibnamefont {Piilo}},\ and\ \bibinfo {author} {\bibfnamefont {B.}~\bibnamefont {Vacchini}},\ }\bibfield  {title} {\bibinfo {title} {Colloquium: Non-markovian dynamics in open quantum systems},\ }\href@noop {} {\bibfield  {journal} {\bibinfo  {journal} {Reviews of Modern Physics}\ }\textbf {\bibinfo {volume} {88}},\ \bibinfo {pages} {021002} (\bibinfo {year} {2016})}\BibitemShut {NoStop}%
\bibitem [{\citenamefont {Duan}\ \emph {et~al.}(2000)\citenamefont {Duan}, \citenamefont {Cirac}, \citenamefont {Zoller},\ and\ \citenamefont {Polzik}}]{duan2000quantum}%
  \BibitemOpen
  \bibfield  {author} {\bibinfo {author} {\bibfnamefont {L.-M.}\ \bibnamefont {Duan}}, \bibinfo {author} {\bibfnamefont {J.}~\bibnamefont {Cirac}}, \bibinfo {author} {\bibfnamefont {P.}~\bibnamefont {Zoller}},\ and\ \bibinfo {author} {\bibfnamefont {E.}~\bibnamefont {Polzik}},\ }\bibfield  {title} {\bibinfo {title} {Quantum communication between atomic ensembles using coherent light},\ }\href@noop {} {\bibfield  {journal} {\bibinfo  {journal} {Physical Review Letters}\ }\textbf {\bibinfo {volume} {85}},\ \bibinfo {pages} {5643} (\bibinfo {year} {2000})}\BibitemShut {NoStop}%
\bibitem [{\citenamefont {Gardiner}\ and\ \citenamefont {Zoller}(2004)}]{gardiner2004quantum}%
  \BibitemOpen
  \bibfield  {author} {\bibinfo {author} {\bibfnamefont {C.}~\bibnamefont {Gardiner}}\ and\ \bibinfo {author} {\bibfnamefont {P.}~\bibnamefont {Zoller}},\ }\href@noop {} {\emph {\bibinfo {title} {Quantum noise: a handbook of Markovian and non-Markovian quantum stochastic methods with applications to quantum optics}}}\ (\bibinfo  {publisher} {Springer Science \& Business Media},\ \bibinfo {year} {2004})\BibitemShut {NoStop}%
\bibitem [{\citenamefont {Kraus}\ \emph {et~al.}(2008)\citenamefont {Kraus}, \citenamefont {B{\"u}chler}, \citenamefont {Diehl}, \citenamefont {Kantian}, \citenamefont {Micheli},\ and\ \citenamefont {Zoller}}]{kraus2008preparation}%
  \BibitemOpen
  \bibfield  {author} {\bibinfo {author} {\bibfnamefont {B.}~\bibnamefont {Kraus}}, \bibinfo {author} {\bibfnamefont {H.~P.}\ \bibnamefont {B{\"u}chler}}, \bibinfo {author} {\bibfnamefont {S.}~\bibnamefont {Diehl}}, \bibinfo {author} {\bibfnamefont {A.}~\bibnamefont {Kantian}}, \bibinfo {author} {\bibfnamefont {A.}~\bibnamefont {Micheli}},\ and\ \bibinfo {author} {\bibfnamefont {P.}~\bibnamefont {Zoller}},\ }\bibfield  {title} {\bibinfo {title} {Preparation of entangled states by quantum markov processes},\ }\href@noop {} {\bibfield  {journal} {\bibinfo  {journal} {Physical Review A—Atomic, Molecular, and Optical Physics}\ }\textbf {\bibinfo {volume} {78}},\ \bibinfo {pages} {042307} (\bibinfo {year} {2008})}\BibitemShut {NoStop}%
\bibitem [{\citenamefont {Lindblad}(1976)}]{lindblad1976generators}%
  \BibitemOpen
  \bibfield  {author} {\bibinfo {author} {\bibfnamefont {G.}~\bibnamefont {Lindblad}},\ }\bibfield  {title} {\bibinfo {title} {On the generators of quantum dynamical semigroups},\ }\href@noop {} {\bibfield  {journal} {\bibinfo  {journal} {Communications in mathematical physics}\ }\textbf {\bibinfo {volume} {48}},\ \bibinfo {pages} {119} (\bibinfo {year} {1976})}\BibitemShut {NoStop}%
\bibitem [{\citenamefont {Gardiner}\ \emph {et~al.}(1992)\citenamefont {Gardiner}, \citenamefont {Parkins},\ and\ \citenamefont {Zoller}}]{gardiner1992wave}%
  \BibitemOpen
  \bibfield  {author} {\bibinfo {author} {\bibfnamefont {C.~W.}\ \bibnamefont {Gardiner}}, \bibinfo {author} {\bibfnamefont {A.~S.}\ \bibnamefont {Parkins}},\ and\ \bibinfo {author} {\bibfnamefont {P.}~\bibnamefont {Zoller}},\ }\bibfield  {title} {\bibinfo {title} {Wave-function quantum stochastic differential equations and quantum-jump simulation methods},\ }\href@noop {} {\bibfield  {journal} {\bibinfo  {journal} {Physical Review A}\ }\textbf {\bibinfo {volume} {46}},\ \bibinfo {pages} {4363} (\bibinfo {year} {1992})}\BibitemShut {NoStop}%
\bibitem [{\citenamefont {Parkins}\ \emph {et~al.}(1993)\citenamefont {Parkins}, \citenamefont {Zoller},\ and\ \citenamefont {Carmichael}}]{parkins1993spectral}%
  \BibitemOpen
  \bibfield  {author} {\bibinfo {author} {\bibfnamefont {A.}~\bibnamefont {Parkins}}, \bibinfo {author} {\bibfnamefont {P.}~\bibnamefont {Zoller}},\ and\ \bibinfo {author} {\bibfnamefont {H.}~\bibnamefont {Carmichael}},\ }\bibfield  {title} {\bibinfo {title} {Spectral linewidth narrowing in a strongly coupled atom-cavity system via squeezed-light excitation of a ‘‘vacuum’’rabi resonance},\ }\href@noop {} {\bibfield  {journal} {\bibinfo  {journal} {Physical Review A}\ }\textbf {\bibinfo {volume} {48}},\ \bibinfo {pages} {758} (\bibinfo {year} {1993})}\BibitemShut {NoStop}%
\bibitem [{\citenamefont {Gardiner}\ and\ \citenamefont {Zoller}(2015)}]{gardiner2015quantum}%
  \BibitemOpen
  \bibfield  {author} {\bibinfo {author} {\bibfnamefont {C.}~\bibnamefont {Gardiner}}\ and\ \bibinfo {author} {\bibfnamefont {P.}~\bibnamefont {Zoller}},\ }\href@noop {} {\emph {\bibinfo {title} {The quantum world of ultra-cold atoms and light book II: the physics of quantum-optical devices}}},\ Vol.~\bibinfo {volume} {4}\ (\bibinfo  {publisher} {World Scientific Publishing Company},\ \bibinfo {year} {2015})\BibitemShut {NoStop}%
\bibitem [{\citenamefont {Ulhaq}\ \emph {et~al.}(2012)\citenamefont {Ulhaq}, \citenamefont {Weiler}, \citenamefont {Ulrich}, \citenamefont {Ro{\ss}bach}, \citenamefont {Jetter},\ and\ \citenamefont {Michler}}]{ulhaq2012cascaded}%
  \BibitemOpen
  \bibfield  {author} {\bibinfo {author} {\bibfnamefont {A.}~\bibnamefont {Ulhaq}}, \bibinfo {author} {\bibfnamefont {S.}~\bibnamefont {Weiler}}, \bibinfo {author} {\bibfnamefont {S.~M.}\ \bibnamefont {Ulrich}}, \bibinfo {author} {\bibfnamefont {R.}~\bibnamefont {Ro{\ss}bach}}, \bibinfo {author} {\bibfnamefont {M.}~\bibnamefont {Jetter}},\ and\ \bibinfo {author} {\bibfnamefont {P.}~\bibnamefont {Michler}},\ }\bibfield  {title} {\bibinfo {title} {Cascaded single-photon emission from the {M}ollow triplet sidebands of a quantum dot},\ }\href@noop {} {\bibfield  {journal} {\bibinfo  {journal} {Nature Photonics}\ }\textbf {\bibinfo {volume} {6}},\ \bibinfo {pages} {238} (\bibinfo {year} {2012})}\BibitemShut {NoStop}%
\bibitem [{\citenamefont {Clark}\ \emph {et~al.}(2003)\citenamefont {Clark}, \citenamefont {Peng}, \citenamefont {Gu},\ and\ \citenamefont {Parkins}}]{clark2003unconditional}%
  \BibitemOpen
  \bibfield  {author} {\bibinfo {author} {\bibfnamefont {S.}~\bibnamefont {Clark}}, \bibinfo {author} {\bibfnamefont {A.}~\bibnamefont {Peng}}, \bibinfo {author} {\bibfnamefont {M.}~\bibnamefont {Gu}},\ and\ \bibinfo {author} {\bibfnamefont {S.}~\bibnamefont {Parkins}},\ }\bibfield  {title} {\bibinfo {title} {Unconditional preparation of entanglement between atoms in cascaded optical cavities},\ }\href@noop {} {\bibfield  {journal} {\bibinfo  {journal} {Physical Review Letters}\ }\textbf {\bibinfo {volume} {91}},\ \bibinfo {pages} {177901} (\bibinfo {year} {2003})}\BibitemShut {NoStop}%
\bibitem [{\citenamefont {Liedl}\ \emph {et~al.}(2024)\citenamefont {Liedl}, \citenamefont {Tebbenjohanns}, \citenamefont {Bach}, \citenamefont {Pucher}, \citenamefont {Rauschenbeutel},\ and\ \citenamefont {Schneeweiss}}]{liedl2024observation}%
  \BibitemOpen
  \bibfield  {author} {\bibinfo {author} {\bibfnamefont {C.}~\bibnamefont {Liedl}}, \bibinfo {author} {\bibfnamefont {F.}~\bibnamefont {Tebbenjohanns}}, \bibinfo {author} {\bibfnamefont {C.}~\bibnamefont {Bach}}, \bibinfo {author} {\bibfnamefont {S.}~\bibnamefont {Pucher}}, \bibinfo {author} {\bibfnamefont {A.}~\bibnamefont {Rauschenbeutel}},\ and\ \bibinfo {author} {\bibfnamefont {P.}~\bibnamefont {Schneeweiss}},\ }\bibfield  {title} {\bibinfo {title} {Observation of superradiant bursts in a cascaded quantum system},\ }\href@noop {} {\bibfield  {journal} {\bibinfo  {journal} {Physical Review X}\ }\textbf {\bibinfo {volume} {14}},\ \bibinfo {pages} {011020} (\bibinfo {year} {2024})}\BibitemShut {NoStop}%
\bibitem [{\citenamefont {Su{\'a}rez-Forero}\ \emph {et~al.}(2025)\citenamefont {Su{\'a}rez-Forero}, \citenamefont {Jalali~Mehrabad}, \citenamefont {Vega}, \citenamefont {Gonz{\'a}lez-Tudela},\ and\ \citenamefont {Hafezi}}]{suarez2025chiral}%
  \BibitemOpen
  \bibfield  {author} {\bibinfo {author} {\bibfnamefont {D.}~\bibnamefont {Su{\'a}rez-Forero}}, \bibinfo {author} {\bibfnamefont {M.}~\bibnamefont {Jalali~Mehrabad}}, \bibinfo {author} {\bibfnamefont {C.}~\bibnamefont {Vega}}, \bibinfo {author} {\bibfnamefont {A.}~\bibnamefont {Gonz{\'a}lez-Tudela}},\ and\ \bibinfo {author} {\bibfnamefont {M.}~\bibnamefont {Hafezi}},\ }\bibfield  {title} {\bibinfo {title} {Chiral quantum optics: Recent developments and future directions},\ }\href@noop {} {\bibfield  {journal} {\bibinfo  {journal} {PRX Quantum}\ }\textbf {\bibinfo {volume} {6}},\ \bibinfo {pages} {020101} (\bibinfo {year} {2025})}\BibitemShut {NoStop}%
\bibitem [{\citenamefont {Wiseman}\ and\ \citenamefont {Milburn}(1993)}]{wiseman1993quantum}%
  \BibitemOpen
  \bibfield  {author} {\bibinfo {author} {\bibfnamefont {H.~M.}\ \bibnamefont {Wiseman}}\ and\ \bibinfo {author} {\bibfnamefont {G.~J.}\ \bibnamefont {Milburn}},\ }\bibfield  {title} {\bibinfo {title} {Quantum theory of optical feedback via homodyne detection},\ }\href@noop {} {\bibfield  {journal} {\bibinfo  {journal} {Physical Review Letters}\ }\textbf {\bibinfo {volume} {70}},\ \bibinfo {pages} {548} (\bibinfo {year} {1993})}\BibitemShut {NoStop}%
\bibitem [{\citenamefont {Whalen}\ \emph {et~al.}(2017)\citenamefont {Whalen}, \citenamefont {Grimsmo},\ and\ \citenamefont {Carmichael}}]{whalen2017open}%
  \BibitemOpen
  \bibfield  {author} {\bibinfo {author} {\bibfnamefont {S.}~\bibnamefont {Whalen}}, \bibinfo {author} {\bibfnamefont {A.}~\bibnamefont {Grimsmo}},\ and\ \bibinfo {author} {\bibfnamefont {H.}~\bibnamefont {Carmichael}},\ }\bibfield  {title} {\bibinfo {title} {Open quantum systems with delayed coherent feedback},\ }\href@noop {} {\bibfield  {journal} {\bibinfo  {journal} {Quantum Science and Technology}\ }\textbf {\bibinfo {volume} {2}},\ \bibinfo {pages} {044008} (\bibinfo {year} {2017})}\BibitemShut {NoStop}%
\bibitem [{\citenamefont {N{\'e}met}\ and\ \citenamefont {Parkins}(2016)}]{nemet2016enhanced}%
  \BibitemOpen
  \bibfield  {author} {\bibinfo {author} {\bibfnamefont {N.}~\bibnamefont {N{\'e}met}}\ and\ \bibinfo {author} {\bibfnamefont {S.}~\bibnamefont {Parkins}},\ }\bibfield  {title} {\bibinfo {title} {Enhanced optical squeezing from a degenerate parametric amplifier via time-delayed coherent feedback},\ }\href@noop {} {\bibfield  {journal} {\bibinfo  {journal} {Physical Review A}\ }\textbf {\bibinfo {volume} {94}},\ \bibinfo {pages} {023809} (\bibinfo {year} {2016})}\BibitemShut {NoStop}%
\bibitem [{\citenamefont {N{\'e}met}\ \emph {et~al.}(2019)\citenamefont {N{\'e}met}, \citenamefont {Parkins}, \citenamefont {Knorr},\ and\ \citenamefont {Carmele}}]{nemet2019stabilizing}%
  \BibitemOpen
  \bibfield  {author} {\bibinfo {author} {\bibfnamefont {N.}~\bibnamefont {N{\'e}met}}, \bibinfo {author} {\bibfnamefont {S.}~\bibnamefont {Parkins}}, \bibinfo {author} {\bibfnamefont {A.}~\bibnamefont {Knorr}},\ and\ \bibinfo {author} {\bibfnamefont {A.}~\bibnamefont {Carmele}},\ }\bibfield  {title} {\bibinfo {title} {Stabilizing quantum coherence against pure dephasing in the presence of time-delayed coherent feedback at finite temperature},\ }\href@noop {} {\bibfield  {journal} {\bibinfo  {journal} {Physical Review A}\ }\textbf {\bibinfo {volume} {99}},\ \bibinfo {pages} {053809} (\bibinfo {year} {2019})}\BibitemShut {NoStop}%
\bibitem [{\citenamefont {Lambropoulos}\ \emph {et~al.}(2000)\citenamefont {Lambropoulos}, \citenamefont {Nikolopoulos}, \citenamefont {Nielsen},\ and\ \citenamefont {Bay}}]{lambropoulos2000fundamental}%
  \BibitemOpen
  \bibfield  {author} {\bibinfo {author} {\bibfnamefont {P.}~\bibnamefont {Lambropoulos}}, \bibinfo {author} {\bibfnamefont {G.~M.}\ \bibnamefont {Nikolopoulos}}, \bibinfo {author} {\bibfnamefont {T.~R.}\ \bibnamefont {Nielsen}},\ and\ \bibinfo {author} {\bibfnamefont {S.}~\bibnamefont {Bay}},\ }\bibfield  {title} {\bibinfo {title} {Fundamental quantum optics in structured reservoirs},\ }\href@noop {} {\bibfield  {journal} {\bibinfo  {journal} {Reports on Progress in Physics}\ }\textbf {\bibinfo {volume} {63}},\ \bibinfo {pages} {455} (\bibinfo {year} {2000})}\BibitemShut {NoStop}%
\bibitem [{\citenamefont {De~Vega}\ and\ \citenamefont {Alonso}(2017)}]{de2017dynamics}%
  \BibitemOpen
  \bibfield  {author} {\bibinfo {author} {\bibfnamefont {I.}~\bibnamefont {De~Vega}}\ and\ \bibinfo {author} {\bibfnamefont {D.}~\bibnamefont {Alonso}},\ }\bibfield  {title} {\bibinfo {title} {Dynamics of non-markovian open quantum systems},\ }\href@noop {} {\bibfield  {journal} {\bibinfo  {journal} {Reviews of Modern Physics}\ }\textbf {\bibinfo {volume} {89}},\ \bibinfo {pages} {015001} (\bibinfo {year} {2017})}\BibitemShut {NoStop}%
\bibitem [{\citenamefont {Eschner}\ \emph {et~al.}(2001)\citenamefont {Eschner}, \citenamefont {Raab}, \citenamefont {Schmidt-Kaler},\ and\ \citenamefont {Blatt}}]{eschner2001light}%
  \BibitemOpen
  \bibfield  {author} {\bibinfo {author} {\bibfnamefont {J.}~\bibnamefont {Eschner}}, \bibinfo {author} {\bibfnamefont {C.}~\bibnamefont {Raab}}, \bibinfo {author} {\bibfnamefont {F.}~\bibnamefont {Schmidt-Kaler}},\ and\ \bibinfo {author} {\bibfnamefont {R.}~\bibnamefont {Blatt}},\ }\bibfield  {title} {\bibinfo {title} {Light interference from single atoms and their mirror images},\ }\href@noop {} {\bibfield  {journal} {\bibinfo  {journal} {Nature}\ }\textbf {\bibinfo {volume} {413}},\ \bibinfo {pages} {495} (\bibinfo {year} {2001})}\BibitemShut {NoStop}%
\bibitem [{\citenamefont {Dubin}\ \emph {et~al.}(2007)\citenamefont {Dubin}, \citenamefont {Rotter}, \citenamefont {Mukherjee}, \citenamefont {Russo}, \citenamefont {Eschner},\ and\ \citenamefont {Blatt}}]{dubin2007photon}%
  \BibitemOpen
  \bibfield  {author} {\bibinfo {author} {\bibfnamefont {F.}~\bibnamefont {Dubin}}, \bibinfo {author} {\bibfnamefont {D.}~\bibnamefont {Rotter}}, \bibinfo {author} {\bibfnamefont {M.}~\bibnamefont {Mukherjee}}, \bibinfo {author} {\bibfnamefont {C.}~\bibnamefont {Russo}}, \bibinfo {author} {\bibfnamefont {J.}~\bibnamefont {Eschner}},\ and\ \bibinfo {author} {\bibfnamefont {R.}~\bibnamefont {Blatt}},\ }\bibfield  {title} {\bibinfo {title} {Photon correlation versus interference of single-atom fluorescence in a half-cavity},\ }\href@noop {} {\bibfield  {journal} {\bibinfo  {journal} {Physical Review Letters}\ }\textbf {\bibinfo {volume} {98}},\ \bibinfo {pages} {183003} (\bibinfo {year} {2007})}\BibitemShut {NoStop}%
\bibitem [{\citenamefont {Parkins}\ and\ \citenamefont {Gardiner}(1988)}]{parkins1988effect}%
  \BibitemOpen
  \bibfield  {author} {\bibinfo {author} {\bibfnamefont {A.~S.}\ \bibnamefont {Parkins}}\ and\ \bibinfo {author} {\bibfnamefont {C.}~\bibnamefont {Gardiner}},\ }\bibfield  {title} {\bibinfo {title} {Effect of finite-bandwidth squeezing on inhibition of atomic-phase decays},\ }\href@noop {} {\bibfield  {journal} {\bibinfo  {journal} {Physical Review A}\ }\textbf {\bibinfo {volume} {37}},\ \bibinfo {pages} {3867} (\bibinfo {year} {1988})}\BibitemShut {NoStop}%
\bibitem [{\citenamefont {Nemet}(2019)}]{nemet2019time}%
  \BibitemOpen
  \bibfield  {author} {\bibinfo {author} {\bibfnamefont {N.}~\bibnamefont {Nemet}},\ }\href@noop {} {\bibinfo {title} {Time-delayed coherent feedback control for open quantum systems, {P}h{D} thesis, {U}niversity of {A}uckland}} (\bibinfo {year} {2019})\BibitemShut {NoStop}%
\bibitem [{\citenamefont {Steck}(2003)}]{steck2003cesium}%
  \BibitemOpen
  \bibfield  {author} {\bibinfo {author} {\bibfnamefont {D.~A.}\ \bibnamefont {Steck}},\ }\href@noop {} {\bibinfo {title} {Cesium d line data}} (\bibinfo {year} {2003}),\ \bibinfo {note} {available online at \url{https://steck.us/alkalidata/cesiumnumbers.pdf}}\BibitemShut {NoStop}%
\bibitem [{\citenamefont {Gouraud}\ \emph {et~al.}(2015)\citenamefont {Gouraud}, \citenamefont {Maxein}, \citenamefont {Nicolas}, \citenamefont {Morin},\ and\ \citenamefont {Laurat}}]{gouraud2015demonstration}%
  \BibitemOpen
  \bibfield  {author} {\bibinfo {author} {\bibfnamefont {B.}~\bibnamefont {Gouraud}}, \bibinfo {author} {\bibfnamefont {D.}~\bibnamefont {Maxein}}, \bibinfo {author} {\bibfnamefont {A.}~\bibnamefont {Nicolas}}, \bibinfo {author} {\bibfnamefont {O.}~\bibnamefont {Morin}},\ and\ \bibinfo {author} {\bibfnamefont {J.}~\bibnamefont {Laurat}},\ }\bibfield  {title} {\bibinfo {title} {Demonstration of a memory for tightly guided light in an optical nanofiber},\ }\href@noop {} {\bibfield  {journal} {\bibinfo  {journal} {Physical review letters}\ }\textbf {\bibinfo {volume} {114}},\ \bibinfo {pages} {180503} (\bibinfo {year} {2015})}\BibitemShut {NoStop}%
\bibitem [{\citenamefont {Goban}(2015)}]{goban2015strong}%
  \BibitemOpen
  \bibfield  {author} {\bibinfo {author} {\bibfnamefont {A.}~\bibnamefont {Goban}},\ }\emph {\bibinfo {title} {Strong atom-light interactions along nanostructures: Transition from free-space to nanophotonic interfaces}},\ \href@noop {} {Ph.D. thesis},\ \bibinfo  {school} {California Institute of Technology} (\bibinfo {year} {2015})\BibitemShut {NoStop}%
\bibitem [{\citenamefont {Nayak}\ \emph {et~al.}(2018)\citenamefont {Nayak}, \citenamefont {Sadgrove}, \citenamefont {Yalla}, \citenamefont {Le~Kien},\ and\ \citenamefont {Hakuta}}]{nayak2018nanofiber}%
  \BibitemOpen
  \bibfield  {author} {\bibinfo {author} {\bibfnamefont {K.~P.}\ \bibnamefont {Nayak}}, \bibinfo {author} {\bibfnamefont {M.}~\bibnamefont {Sadgrove}}, \bibinfo {author} {\bibfnamefont {R.}~\bibnamefont {Yalla}}, \bibinfo {author} {\bibfnamefont {F.}~\bibnamefont {Le~Kien}},\ and\ \bibinfo {author} {\bibfnamefont {K.}~\bibnamefont {Hakuta}},\ }\bibfield  {title} {\bibinfo {title} {Nanofiber quantum photonics},\ }\href@noop {} {\bibfield  {journal} {\bibinfo  {journal} {Journal of Optics}\ }\textbf {\bibinfo {volume} {20}},\ \bibinfo {pages} {073001} (\bibinfo {year} {2018})}\BibitemShut {NoStop}%
\bibitem [{\citenamefont {Ruddell}\ \emph {et~al.}(2017)\citenamefont {Ruddell}, \citenamefont {Webb}, \citenamefont {Herrera}, \citenamefont {Parkins},\ and\ \citenamefont {Hoogerland}}]{ruddell2017collective}%
  \BibitemOpen
  \bibfield  {author} {\bibinfo {author} {\bibfnamefont {S.~K.}\ \bibnamefont {Ruddell}}, \bibinfo {author} {\bibfnamefont {K.~E.}\ \bibnamefont {Webb}}, \bibinfo {author} {\bibfnamefont {I.}~\bibnamefont {Herrera}}, \bibinfo {author} {\bibfnamefont {A.~S.}\ \bibnamefont {Parkins}},\ and\ \bibinfo {author} {\bibfnamefont {M.~D.}\ \bibnamefont {Hoogerland}},\ }\bibfield  {title} {\bibinfo {title} {Collective strong coupling of cold atoms to an all-fiber ring cavity},\ }\href@noop {} {\bibfield  {journal} {\bibinfo  {journal} {Optica}\ }\textbf {\bibinfo {volume} {4}},\ \bibinfo {pages} {576} (\bibinfo {year} {2017})}\BibitemShut {NoStop}%
\bibitem [{\citenamefont {Lechner}\ \emph {et~al.}(2023)\citenamefont {Lechner}, \citenamefont {Pennetta}, \citenamefont {Blaha}, \citenamefont {Schneeweiss}, \citenamefont {Rauschenbeutel},\ and\ \citenamefont {Volz}}]{lechner2023light}%
  \BibitemOpen
  \bibfield  {author} {\bibinfo {author} {\bibfnamefont {D.}~\bibnamefont {Lechner}}, \bibinfo {author} {\bibfnamefont {R.}~\bibnamefont {Pennetta}}, \bibinfo {author} {\bibfnamefont {M.}~\bibnamefont {Blaha}}, \bibinfo {author} {\bibfnamefont {P.}~\bibnamefont {Schneeweiss}}, \bibinfo {author} {\bibfnamefont {A.}~\bibnamefont {Rauschenbeutel}},\ and\ \bibinfo {author} {\bibfnamefont {J.}~\bibnamefont {Volz}},\ }\bibfield  {title} {\bibinfo {title} {Light-matter interaction at the transition between cavity and waveguide {QED}},\ }\href@noop {} {\bibfield  {journal} {\bibinfo  {journal} {preprint arXiv:2302.07161}\ } (\bibinfo {year} {2023})}\BibitemShut {NoStop}%
\bibitem [{\citenamefont {Solano}\ \emph {et~al.}(2017)\citenamefont {Solano}, \citenamefont {Barberis-Blostein}, \citenamefont {Fatemi}, \citenamefont {Orozco},\ and\ \citenamefont {Rolston}}]{solano2017super}%
  \BibitemOpen
  \bibfield  {author} {\bibinfo {author} {\bibfnamefont {P.}~\bibnamefont {Solano}}, \bibinfo {author} {\bibfnamefont {P.}~\bibnamefont {Barberis-Blostein}}, \bibinfo {author} {\bibfnamefont {F.~K.}\ \bibnamefont {Fatemi}}, \bibinfo {author} {\bibfnamefont {L.~A.}\ \bibnamefont {Orozco}},\ and\ \bibinfo {author} {\bibfnamefont {S.~L.}\ \bibnamefont {Rolston}},\ }\bibfield  {title} {\bibinfo {title} {Super-radiance reveals infinite-range dipole interactions through a nanofiber},\ }\href@noop {} {\bibfield  {journal} {\bibinfo  {journal} {Nature Communications}\ }\textbf {\bibinfo {volume} {8}},\ \bibinfo {pages} {1857} (\bibinfo {year} {2017})}\BibitemShut {NoStop}%
\bibitem [{\citenamefont {Corzo}\ \emph {et~al.}(2019)\citenamefont {Corzo}, \citenamefont {Raskop}, \citenamefont {Chandra}, \citenamefont {Sheremet}, \citenamefont {Gouraud},\ and\ \citenamefont {Laurat}}]{corzo2019waveguide}%
  \BibitemOpen
  \bibfield  {author} {\bibinfo {author} {\bibfnamefont {N.~V.}\ \bibnamefont {Corzo}}, \bibinfo {author} {\bibfnamefont {J.}~\bibnamefont {Raskop}}, \bibinfo {author} {\bibfnamefont {A.}~\bibnamefont {Chandra}}, \bibinfo {author} {\bibfnamefont {A.~S.}\ \bibnamefont {Sheremet}}, \bibinfo {author} {\bibfnamefont {B.}~\bibnamefont {Gouraud}},\ and\ \bibinfo {author} {\bibfnamefont {J.}~\bibnamefont {Laurat}},\ }\bibfield  {title} {\bibinfo {title} {Waveguide-coupled single collective excitation of atomic arrays},\ }\href@noop {} {\bibfield  {journal} {\bibinfo  {journal} {Nature}\ }\textbf {\bibinfo {volume} {566}},\ \bibinfo {pages} {359} (\bibinfo {year} {2019})}\BibitemShut {NoStop}%
\bibitem [{\citenamefont {Solano}\ \emph {et~al.}(2019)\citenamefont {Solano}, \citenamefont {Grover}, \citenamefont {Xu}, \citenamefont {Barberis-Blostein}, \citenamefont {Munday}, \citenamefont {Orozco}, \citenamefont {Phillips},\ and\ \citenamefont {Rolston}}]{solano2019alignment}%
  \BibitemOpen
  \bibfield  {author} {\bibinfo {author} {\bibfnamefont {P.}~\bibnamefont {Solano}}, \bibinfo {author} {\bibfnamefont {J.~A.}\ \bibnamefont {Grover}}, \bibinfo {author} {\bibfnamefont {Y.}~\bibnamefont {Xu}}, \bibinfo {author} {\bibfnamefont {P.}~\bibnamefont {Barberis-Blostein}}, \bibinfo {author} {\bibfnamefont {J.~N.}\ \bibnamefont {Munday}}, \bibinfo {author} {\bibfnamefont {L.~A.}\ \bibnamefont {Orozco}}, \bibinfo {author} {\bibfnamefont {W.~D.}\ \bibnamefont {Phillips}},\ and\ \bibinfo {author} {\bibfnamefont {S.~L.}\ \bibnamefont {Rolston}},\ }\bibfield  {title} {\bibinfo {title} {Alignment-dependent decay rate of an atomic dipole near an optical nanofiber},\ }\href@noop {} {\bibfield  {journal} {\bibinfo  {journal} {Physical Review A}\ }\textbf {\bibinfo {volume} {99}},\ \bibinfo {pages} {013822} (\bibinfo {year} {2019})}\BibitemShut {NoStop}%
\bibitem [{\citenamefont {Nayak}\ \emph {et~al.}(2007)\citenamefont {Nayak}, \citenamefont {Melentiev}, \citenamefont {Morinaga}, \citenamefont {Le~Kien}, \citenamefont {Balykin},\ and\ \citenamefont {Hakuta}}]{nayak2007optical}%
  \BibitemOpen
  \bibfield  {author} {\bibinfo {author} {\bibfnamefont {K.~P.}\ \bibnamefont {Nayak}}, \bibinfo {author} {\bibfnamefont {P.~N.}\ \bibnamefont {Melentiev}}, \bibinfo {author} {\bibfnamefont {M.}~\bibnamefont {Morinaga}}, \bibinfo {author} {\bibfnamefont {F.}~\bibnamefont {Le~Kien}}, \bibinfo {author} {\bibfnamefont {V.~I.}\ \bibnamefont {Balykin}},\ and\ \bibinfo {author} {\bibfnamefont {K.}~\bibnamefont {Hakuta}},\ }\bibfield  {title} {\bibinfo {title} {Optical nanofiber as an efficient tool for manipulating and probing atomic fluorescence},\ }\href@noop {} {\bibfield  {journal} {\bibinfo  {journal} {Optics Express}\ }\textbf {\bibinfo {volume} {15}},\ \bibinfo {pages} {5431} (\bibinfo {year} {2007})}\BibitemShut {NoStop}%
\bibitem [{\citenamefont {Das}\ \emph {et~al.}(2010)\citenamefont {Das}, \citenamefont {Shirasaki}, \citenamefont {Nayak}, \citenamefont {Morinaga}, \citenamefont {Le~Kien},\ and\ \citenamefont {Hakuta}}]{das2010measurement}%
  \BibitemOpen
  \bibfield  {author} {\bibinfo {author} {\bibfnamefont {M.}~\bibnamefont {Das}}, \bibinfo {author} {\bibfnamefont {A.}~\bibnamefont {Shirasaki}}, \bibinfo {author} {\bibfnamefont {K.}~\bibnamefont {Nayak}}, \bibinfo {author} {\bibfnamefont {M.}~\bibnamefont {Morinaga}}, \bibinfo {author} {\bibfnamefont {F.}~\bibnamefont {Le~Kien}},\ and\ \bibinfo {author} {\bibfnamefont {K.}~\bibnamefont {Hakuta}},\ }\bibfield  {title} {\bibinfo {title} {Measurement of fluorescence emission spectrum of few strongly driven atoms using an optical nanofiber},\ }\href@noop {} {\bibfield  {journal} {\bibinfo  {journal} {Optics Express}\ }\textbf {\bibinfo {volume} {18}},\ \bibinfo {pages} {17154} (\bibinfo {year} {2010})}\BibitemShut {NoStop}%
\bibitem [{\citenamefont {Sinha}\ \emph {et~al.}(2020)\citenamefont {Sinha}, \citenamefont {Meystre}, \citenamefont {Goldschmidt}, \citenamefont {Fatemi}, \citenamefont {Rolston},\ and\ \citenamefont {Solano}}]{sinha2020non}%
  \BibitemOpen
  \bibfield  {author} {\bibinfo {author} {\bibfnamefont {K.}~\bibnamefont {Sinha}}, \bibinfo {author} {\bibfnamefont {P.}~\bibnamefont {Meystre}}, \bibinfo {author} {\bibfnamefont {E.~A.}\ \bibnamefont {Goldschmidt}}, \bibinfo {author} {\bibfnamefont {F.~K.}\ \bibnamefont {Fatemi}}, \bibinfo {author} {\bibfnamefont {S.~L.}\ \bibnamefont {Rolston}},\ and\ \bibinfo {author} {\bibfnamefont {P.}~\bibnamefont {Solano}},\ }\bibfield  {title} {\bibinfo {title} {Non-markovian collective emission from macroscopically separated emitters},\ }\href@noop {} {\bibfield  {journal} {\bibinfo  {journal} {Physical Review Letters}\ }\textbf {\bibinfo {volume} {124}},\ \bibinfo {pages} {043603} (\bibinfo {year} {2020})}\BibitemShut {NoStop}%
\bibitem [{\citenamefont {Ferreira}\ \emph {et~al.}(2021)\citenamefont {Ferreira}, \citenamefont {Banker}, \citenamefont {Sipahigil}, \citenamefont {Matheny}, \citenamefont {Keller}, \citenamefont {Kim}, \citenamefont {Mirhosseini},\ and\ \citenamefont {Painter}}]{ferreira2021collapse}%
  \BibitemOpen
  \bibfield  {author} {\bibinfo {author} {\bibfnamefont {V.~S.}\ \bibnamefont {Ferreira}}, \bibinfo {author} {\bibfnamefont {J.}~\bibnamefont {Banker}}, \bibinfo {author} {\bibfnamefont {A.}~\bibnamefont {Sipahigil}}, \bibinfo {author} {\bibfnamefont {M.~H.}\ \bibnamefont {Matheny}}, \bibinfo {author} {\bibfnamefont {A.~J.}\ \bibnamefont {Keller}}, \bibinfo {author} {\bibfnamefont {E.}~\bibnamefont {Kim}}, \bibinfo {author} {\bibfnamefont {M.}~\bibnamefont {Mirhosseini}},\ and\ \bibinfo {author} {\bibfnamefont {O.}~\bibnamefont {Painter}},\ }\bibfield  {title} {\bibinfo {title} {Collapse and revival of an artificial atom coupled to a structured photonic reservoir},\ }\href@noop {} {\bibfield  {journal} {\bibinfo  {journal} {Physical Review X}\ }\textbf {\bibinfo {volume} {11}},\ \bibinfo {pages} {041043} (\bibinfo {year} {2021})}\BibitemShut {NoStop}%
\bibitem [{\citenamefont {Carmichael}(2007)}]{carmichael2007statistical}%
  \BibitemOpen
  \bibfield  {author} {\bibinfo {author} {\bibfnamefont {H.~J.}\ \bibnamefont {Carmichael}},\ }\href@noop {} {\emph {\bibinfo {title} {Statistical methods in quantum optics 2: Non-classical fields}}}\ (\bibinfo  {publisher} {Springer Science \& Business Media},\ \bibinfo {year} {2007})\BibitemShut {NoStop}%
\bibitem [{\citenamefont {Chu}(1991)}]{chu1991laser}%
  \BibitemOpen
  \bibfield  {author} {\bibinfo {author} {\bibfnamefont {S.}~\bibnamefont {Chu}},\ }\bibfield  {title} {\bibinfo {title} {Laser manipulation of atoms and particles},\ }\href@noop {} {\bibfield  {journal} {\bibinfo  {journal} {Science}\ }\textbf {\bibinfo {volume} {253}},\ \bibinfo {pages} {861} (\bibinfo {year} {1991})}\BibitemShut {NoStop}%
\bibitem [{\citenamefont {Ruddell}(2017)}]{ruddell2017calorimetry}%
  \BibitemOpen
  \bibfield  {author} {\bibinfo {author} {\bibfnamefont {S.}~\bibnamefont {Ruddell}},\ }\emph {\bibinfo {title} {Calorimetry of an ultracold {B}ose gas and cavity quantum electrodynamics with an optical nanofibre}},\ \href@noop {} {Ph.D. thesis},\ \bibinfo  {school} {University of Auckland} (\bibinfo {year} {2017})\BibitemShut {NoStop}%
\bibitem [{\citenamefont {Birks}\ and\ \citenamefont {Li}(1992)}]{birks1992shape}%
  \BibitemOpen
  \bibfield  {author} {\bibinfo {author} {\bibfnamefont {T.~A.}\ \bibnamefont {Birks}}\ and\ \bibinfo {author} {\bibfnamefont {Y.~W.}\ \bibnamefont {Li}},\ }\bibfield  {title} {\bibinfo {title} {The shape of fiber tapers},\ }\href@noop {} {\bibfield  {journal} {\bibinfo  {journal} {Journal of lightwave technology}\ }\textbf {\bibinfo {volume} {10}},\ \bibinfo {pages} {432} (\bibinfo {year} {1992})}\BibitemShut {NoStop}%
\bibitem [{\citenamefont {Karapetyan}\ \emph {et~al.}(2011)\citenamefont {Karapetyan}, \citenamefont {Alt},\ and\ \citenamefont {Meshchede}}]{karapetyan2011optical}%
  \BibitemOpen
  \bibfield  {author} {\bibinfo {author} {\bibfnamefont {K.}~\bibnamefont {Karapetyan}}, \bibinfo {author} {\bibfnamefont {W.}~\bibnamefont {Alt}},\ and\ \bibinfo {author} {\bibfnamefont {D.}~\bibnamefont {Meshchede}},\ }\href@noop {} {\bibinfo {title} {Optical fibre toolbox for matlab, version 2.1}} (\bibinfo {year} {2011})\BibitemShut {NoStop}%
\bibitem [{\citenamefont {Nagai}\ and\ \citenamefont {Aoki}(2014)}]{nagai2014ultra}%
  \BibitemOpen
  \bibfield  {author} {\bibinfo {author} {\bibfnamefont {R.}~\bibnamefont {Nagai}}\ and\ \bibinfo {author} {\bibfnamefont {T.}~\bibnamefont {Aoki}},\ }\bibfield  {title} {\bibinfo {title} {Ultra-low-loss tapered optical fibers with minimal lengths},\ }\href@noop {} {\bibfield  {journal} {\bibinfo  {journal} {Optics express}\ }\textbf {\bibinfo {volume} {22}},\ \bibinfo {pages} {28427} (\bibinfo {year} {2014})}\BibitemShut {NoStop}%
\bibitem [{\citenamefont {Nayak}\ \emph {et~al.}(2012)\citenamefont {Nayak}, \citenamefont {Das}, \citenamefont {Le~Kien},\ and\ \citenamefont {Hakuta}}]{nayak2012spectroscopy}%
  \BibitemOpen
  \bibfield  {author} {\bibinfo {author} {\bibfnamefont {K.}~\bibnamefont {Nayak}}, \bibinfo {author} {\bibfnamefont {M.}~\bibnamefont {Das}}, \bibinfo {author} {\bibfnamefont {F.}~\bibnamefont {Le~Kien}},\ and\ \bibinfo {author} {\bibfnamefont {K.}~\bibnamefont {Hakuta}},\ }\bibfield  {title} {\bibinfo {title} {Spectroscopy of near-surface atoms using an optical nanofiber},\ }\href@noop {} {\bibfield  {journal} {\bibinfo  {journal} {Optics Communications}\ }\textbf {\bibinfo {volume} {285}},\ \bibinfo {pages} {4698} (\bibinfo {year} {2012})}\BibitemShut {NoStop}%
\bibitem [{\citenamefont {Passerat~de Silans}\ \emph {et~al.}(2006)\citenamefont {Passerat~de Silans}, \citenamefont {Farias}, \citenamefont {Ori{\'a}},\ and\ \citenamefont {Chevrollier}}]{passerat2006laser}%
  \BibitemOpen
  \bibfield  {author} {\bibinfo {author} {\bibfnamefont {T.}~\bibnamefont {Passerat~de Silans}}, \bibinfo {author} {\bibfnamefont {B.}~\bibnamefont {Farias}}, \bibinfo {author} {\bibfnamefont {M.}~\bibnamefont {Ori{\'a}}},\ and\ \bibinfo {author} {\bibfnamefont {M.}~\bibnamefont {Chevrollier}},\ }\bibfield  {title} {\bibinfo {title} {Laser-induced quantum adsorption of neutral atoms in dielectric surfaces},\ }\href@noop {} {\bibfield  {journal} {\bibinfo  {journal} {Applied Physics B}\ }\textbf {\bibinfo {volume} {82}},\ \bibinfo {pages} {367} (\bibinfo {year} {2006})}\BibitemShut {NoStop}%
\bibitem [{\citenamefont {Kien}\ \emph {et~al.}(2007{\natexlab{a}})\citenamefont {Kien}, \citenamefont {Dutta~Gupta},\ and\ \citenamefont {Hakuta}}]{kien2007phonon}%
  \BibitemOpen
  \bibfield  {author} {\bibinfo {author} {\bibfnamefont {F.~L.}\ \bibnamefont {Kien}}, \bibinfo {author} {\bibfnamefont {S.}~\bibnamefont {Dutta~Gupta}},\ and\ \bibinfo {author} {\bibfnamefont {K.}~\bibnamefont {Hakuta}},\ }\bibfield  {title} {\bibinfo {title} {Phonon-mediated decay of an atom in a surface-induced potential},\ }\href@noop {} {\bibfield  {journal} {\bibinfo  {journal} {Physical Review A—Atomic, Molecular, and Optical Physics}\ }\textbf {\bibinfo {volume} {75}},\ \bibinfo {pages} {062904} (\bibinfo {year} {2007}{\natexlab{a}})}\BibitemShut {NoStop}%
\bibitem [{\citenamefont {Kien}\ \emph {et~al.}(2007{\natexlab{b}})\citenamefont {Kien}, \citenamefont {Gupta},\ and\ \citenamefont {Hakuta}}]{kien2007optical}%
  \BibitemOpen
  \bibfield  {author} {\bibinfo {author} {\bibfnamefont {F.~L.}\ \bibnamefont {Kien}}, \bibinfo {author} {\bibfnamefont {S.~D.}\ \bibnamefont {Gupta}},\ and\ \bibinfo {author} {\bibfnamefont {K.}~\bibnamefont {Hakuta}},\ }\bibfield  {title} {\bibinfo {title} {Optical excitation spectrum of an atom in a surface-induced potential},\ }\href@noop {} {\bibfield  {journal} {\bibinfo  {journal} {Physical Review A—Atomic, Molecular, and Optical Physics}\ }\textbf {\bibinfo {volume} {75}},\ \bibinfo {pages} {032508} (\bibinfo {year} {2007}{\natexlab{b}})}\BibitemShut {NoStop}%
\bibitem [{\citenamefont {Le~Kien}\ and\ \citenamefont {Hakuta}(2007)}]{le2007spontaneous}%
  \BibitemOpen
  \bibfield  {author} {\bibinfo {author} {\bibfnamefont {F.}~\bibnamefont {Le~Kien}}\ and\ \bibinfo {author} {\bibfnamefont {K.}~\bibnamefont {Hakuta}},\ }\bibfield  {title} {\bibinfo {title} {Spontaneous radiative decay of translational levels of an atom near a dielectric surface},\ }\href@noop {} {\bibfield  {journal} {\bibinfo  {journal} {Physical Review A—Atomic, Molecular, and Optical Physics}\ }\textbf {\bibinfo {volume} {75}},\ \bibinfo {pages} {013423} (\bibinfo {year} {2007})}\BibitemShut {NoStop}%
\bibitem [{\citenamefont {Patterson}\ \emph {et~al.}(2018)\citenamefont {Patterson}, \citenamefont {Solano}, \citenamefont {Julienne}, \citenamefont {Orozco},\ and\ \citenamefont {Rolston}}]{patterson2018spectral}%
  \BibitemOpen
  \bibfield  {author} {\bibinfo {author} {\bibfnamefont {B.}~\bibnamefont {Patterson}}, \bibinfo {author} {\bibfnamefont {P.}~\bibnamefont {Solano}}, \bibinfo {author} {\bibfnamefont {P.}~\bibnamefont {Julienne}}, \bibinfo {author} {\bibfnamefont {L.}~\bibnamefont {Orozco}},\ and\ \bibinfo {author} {\bibfnamefont {S.}~\bibnamefont {Rolston}},\ }\bibfield  {title} {\bibinfo {title} {Spectral asymmetry of atoms in the van der waals potential of an optical nanofiber},\ }\href@noop {} {\bibfield  {journal} {\bibinfo  {journal} {Physical Review A}\ }\textbf {\bibinfo {volume} {97}},\ \bibinfo {pages} {032509} (\bibinfo {year} {2018})}\BibitemShut {NoStop}%
\bibitem [{\citenamefont {Metcalf}\ and\ \citenamefont {Van~der Straten}(1999)}]{metcalf1999laser}%
  \BibitemOpen
  \bibfield  {author} {\bibinfo {author} {\bibfnamefont {H.~J.}\ \bibnamefont {Metcalf}}\ and\ \bibinfo {author} {\bibfnamefont {P.}~\bibnamefont {Van~der Straten}},\ }\href@noop {} {\emph {\bibinfo {title} {Laser cooling and trapping}}}\ (\bibinfo  {publisher} {Springer Science \& Business Media},\ \bibinfo {year} {1999})\BibitemShut {NoStop}%
\bibitem [{\citenamefont {Ott}\ \emph {et~al.}(2013)\citenamefont {Ott}, \citenamefont {Wubs}, \citenamefont {Lodahl}, \citenamefont {Mortensen},\ and\ \citenamefont {Kaiser}}]{ott2013cooperative}%
  \BibitemOpen
  \bibfield  {author} {\bibinfo {author} {\bibfnamefont {J.~R.}\ \bibnamefont {Ott}}, \bibinfo {author} {\bibfnamefont {M.}~\bibnamefont {Wubs}}, \bibinfo {author} {\bibfnamefont {P.}~\bibnamefont {Lodahl}}, \bibinfo {author} {\bibfnamefont {N.~A.}\ \bibnamefont {Mortensen}},\ and\ \bibinfo {author} {\bibfnamefont {R.}~\bibnamefont {Kaiser}},\ }\bibfield  {title} {\bibinfo {title} {Cooperative fluorescence from a strongly driven dilute cloud of atoms},\ }\href@noop {} {\bibfield  {journal} {\bibinfo  {journal} {Physical Review A—Atomic, Molecular, and Optical Physics}\ }\textbf {\bibinfo {volume} {87}},\ \bibinfo {pages} {061801} (\bibinfo {year} {2013})}\BibitemShut {NoStop}%
\bibitem [{\citenamefont {Mollow}(1969)}]{mollow1969power}%
  \BibitemOpen
  \bibfield  {author} {\bibinfo {author} {\bibfnamefont {B.~R.}\ \bibnamefont {Mollow}},\ }\bibfield  {title} {\bibinfo {title} {Power spectrum of light scattered by two-level systems},\ }\href@noop {} {\bibfield  {journal} {\bibinfo  {journal} {Physical Review}\ }\textbf {\bibinfo {volume} {188}} (\bibinfo {year} {1969})}\BibitemShut {NoStop}%
\bibitem [{\citenamefont {Ortiz-Guti{\'e}rrez}\ \emph {et~al.}(2019)\citenamefont {Ortiz-Guti{\'e}rrez}, \citenamefont {Teixeira}, \citenamefont {Eloy}, \citenamefont {da~Silva}, \citenamefont {Kaiser}, \citenamefont {Bachelard},\ and\ \citenamefont {Fouch{\'e}}}]{ortiz2019mollow}%
  \BibitemOpen
  \bibfield  {author} {\bibinfo {author} {\bibfnamefont {L.}~\bibnamefont {Ortiz-Guti{\'e}rrez}}, \bibinfo {author} {\bibfnamefont {R.~C.}\ \bibnamefont {Teixeira}}, \bibinfo {author} {\bibfnamefont {A.}~\bibnamefont {Eloy}}, \bibinfo {author} {\bibfnamefont {D.~F.}\ \bibnamefont {da~Silva}}, \bibinfo {author} {\bibfnamefont {R.}~\bibnamefont {Kaiser}}, \bibinfo {author} {\bibfnamefont {R.}~\bibnamefont {Bachelard}},\ and\ \bibinfo {author} {\bibfnamefont {M.}~\bibnamefont {Fouch{\'e}}},\ }\bibfield  {title} {\bibinfo {title} {Mollow triplet in cold atoms},\ }\href@noop {} {\bibfield  {journal} {\bibinfo  {journal} {New Journal of Physics}\ }\textbf {\bibinfo {volume} {21}},\ \bibinfo {pages} {093019} (\bibinfo {year} {2019})}\BibitemShut {NoStop}%
\bibitem [{\citenamefont {Ruddell}\ \emph {et~al.}(2020)\citenamefont {Ruddell}, \citenamefont {Webb}, \citenamefont {Takahata}, \citenamefont {Kato},\ and\ \citenamefont {Aoki}}]{ruddell2020ultra}%
  \BibitemOpen
  \bibfield  {author} {\bibinfo {author} {\bibfnamefont {S.~K.}\ \bibnamefont {Ruddell}}, \bibinfo {author} {\bibfnamefont {K.~E.}\ \bibnamefont {Webb}}, \bibinfo {author} {\bibfnamefont {M.}~\bibnamefont {Takahata}}, \bibinfo {author} {\bibfnamefont {S.}~\bibnamefont {Kato}},\ and\ \bibinfo {author} {\bibfnamefont {T.}~\bibnamefont {Aoki}},\ }\bibfield  {title} {\bibinfo {title} {Ultra-low-loss nanofiber fabry--perot cavities optimized for cavity quantum electrodynamics},\ }\href@noop {} {\bibfield  {journal} {\bibinfo  {journal} {Optics letters}\ }\textbf {\bibinfo {volume} {45}},\ \bibinfo {pages} {4875} (\bibinfo {year} {2020})}\BibitemShut {NoStop}%
\bibitem [{\citenamefont {Dicke}(1954)}]{dicke1954coherence}%
  \BibitemOpen
  \bibfield  {author} {\bibinfo {author} {\bibfnamefont {R.~H.}\ \bibnamefont {Dicke}},\ }\bibfield  {title} {\bibinfo {title} {Coherence in spontaneous radiation processes},\ }\href@noop {} {\bibfield  {journal} {\bibinfo  {journal} {Physical review}\ }\textbf {\bibinfo {volume} {93}},\ \bibinfo {pages} {99} (\bibinfo {year} {1954})}\BibitemShut {NoStop}%
\bibitem [{\citenamefont {Tebbenjohanns}\ \emph {et~al.}(2024)\citenamefont {Tebbenjohanns}, \citenamefont {Mink}, \citenamefont {Bach}, \citenamefont {Rauschenbeutel},\ and\ \citenamefont {Fleischhauer}}]{tebbenjohanns2024predicting}%
  \BibitemOpen
  \bibfield  {author} {\bibinfo {author} {\bibfnamefont {F.}~\bibnamefont {Tebbenjohanns}}, \bibinfo {author} {\bibfnamefont {C.~D.}\ \bibnamefont {Mink}}, \bibinfo {author} {\bibfnamefont {C.}~\bibnamefont {Bach}}, \bibinfo {author} {\bibfnamefont {A.}~\bibnamefont {Rauschenbeutel}},\ and\ \bibinfo {author} {\bibfnamefont {M.}~\bibnamefont {Fleischhauer}},\ }\bibfield  {title} {\bibinfo {title} {Predicting correlations in superradiant emission from a cascaded quantum system},\ }\href@noop {} {\bibfield  {journal} {\bibinfo  {journal} {Physical Review A}\ }\textbf {\bibinfo {volume} {110}},\ \bibinfo {pages} {043713} (\bibinfo {year} {2024})}\BibitemShut {NoStop}%
\bibitem [{\citenamefont {N{\'e}met}\ \emph {et~al.}(2020)\citenamefont {N{\'e}met}, \citenamefont {White}, \citenamefont {Kato}, \citenamefont {Parkins},\ and\ \citenamefont {Aoki}}]{nemet2020transfer}%
  \BibitemOpen
  \bibfield  {author} {\bibinfo {author} {\bibfnamefont {N.}~\bibnamefont {N{\'e}met}}, \bibinfo {author} {\bibfnamefont {D.}~\bibnamefont {White}}, \bibinfo {author} {\bibfnamefont {S.}~\bibnamefont {Kato}}, \bibinfo {author} {\bibfnamefont {S.}~\bibnamefont {Parkins}},\ and\ \bibinfo {author} {\bibfnamefont {T.}~\bibnamefont {Aoki}},\ }\bibfield  {title} {\bibinfo {title} {Transfer-matrix approach to determining the linear response of all-fiber networks of cavity-qed systems},\ }\href@noop {} {\bibfield  {journal} {\bibinfo  {journal} {Physical Review Applied}\ }\textbf {\bibinfo {volume} {13}},\ \bibinfo {pages} {064010} (\bibinfo {year} {2020})}\BibitemShut {NoStop}%
\bibitem [{\citenamefont {Pichler}\ and\ \citenamefont {Zoller}(2016)}]{pichler2016photonic}%
  \BibitemOpen
  \bibfield  {author} {\bibinfo {author} {\bibfnamefont {H.}~\bibnamefont {Pichler}}\ and\ \bibinfo {author} {\bibfnamefont {P.}~\bibnamefont {Zoller}},\ }\bibfield  {title} {\bibinfo {title} {Photonic circuits with time delays and quantum feedback},\ }\href@noop {} {\bibfield  {journal} {\bibinfo  {journal} {Physical Review Letters}\ }\textbf {\bibinfo {volume} {116}},\ \bibinfo {pages} {093601} (\bibinfo {year} {2016})}\BibitemShut {NoStop}%
\end{thebibliography}%

\end{document}